\documentclass[times,twocolumn,final]{elsarticle}
\usepackage{medima}
\usepackage{framed,multirow}

\usepackage{amssymb}
\usepackage{latexsym}

\usepackage{url}
\usepackage{xcolor}

\usepackage{hyperref}

\definecolor{newcolor}{rgb}{.8,.349,.1}

\journal{Medical Image Analysis}

\begin{document}

\verso{Boyeong Woo \textit{et~al.}}

\begin{frontmatter}

\title{Automated anomaly-aware 3D segmentation of bones and cartilages in knee MR images from the Osteoarthritis Initiative}

\author[1]{Boyeong \snm{Woo}\corref{cor1}}
\cortext[cor1]{Corresponding author}
\ead{b.woo@uqconnect.edu.au}
\author[2]{Craig \snm{Engstrom}}
\author[2]{William \snm{Baresic}}
\author[1,3]{Jurgen \snm{Fripp}}
\author[1]{Stuart \snm{Crozier}}
\author[1]{Shekhar S. \snm{Chandra}}

\address[1]{School of Information Technology and Electrical Engineering, The University of Queensland, Australia}
\address[2]{School of Human Movement and Nutrition Sciences, The University of Queensland, Australia}
\address[3]{Australian e-Health Research Centre, Commonwealth Scientific and Industrial Research Organisation, Australia}

%\received{1 May 2013}
%\finalform{10 May 2013}
%\accepted{13 May 2013}
%\availableonline{15 May 2013}
%\communicated{S. Sarkar}

\begin{abstract}
% The abstract should be no longer than 200 words.
In medical image analysis, automated segmentation of multi-component anatomical structures, which often have a spectrum of potential anomalies and pathologies, is a challenging task. In this work, we develop a multi-step approach using U-Net-based neural networks to initially detect anomalies (bone marrow lesions, bone cysts) in the distal femur, proximal tibia and patella from 3D magnetic resonance (MR) images of the knee in individuals with varying grades of osteoarthritis. Subsequently, the extracted data are used for downstream tasks involving semantic segmentation of individual bone and cartilage volumes as well as bone anomalies. For anomaly detection, the U-Net-based models were developed to reconstruct the bone profiles of the femur and tibia in images via inpainting so anomalous bone regions could be replaced with close to normal appearances. The reconstruction error was used to detect bone anomalies. A second anomaly-aware network, which was compared to anomaly-na\"ive segmentation networks, was used to provide a final automated segmentation of the femoral, tibial and patellar bones and cartilages from the knee MR images containing a spectrum of bone anomalies. The anomaly-aware segmentation approach provided up to 58\% reduction in Hausdorff distances for bone segmentations compared to the results from the anomaly-na\"ive segmentation networks. In addition, the anomaly-aware networks were able to detect bone lesions in the MR images with greater sensitivity and specificity (area under the receiver operating characteristic curve [AUC] up to 0.896) compared to the anomaly-na\"ive segmentation networks (AUC up to 0.874).
\end{abstract}


\end{frontmatter}

%\linenumbers


\section{Introduction}
\label{sec:introduction}

Deep learning methods, particularly convolutional neural networks (CNNs), have shown promising results in image recognition tasks and are a rapidly evolving area of research in the field of automated medical image analysis. Common deep learning-based computer vision tasks that have been applied to medical imaging include classification, detection, and segmentation \citep{doi:10.1148/radiol.2018180547}. For example, U-Net, proposed by \citet{RFB15a}, has become a popular CNN model for medical image segmentation. There are now several publicly available medical image databases, such as the Osteoarthritis Initiative \citep{OAI}, which provide the basis for deep learning research in medical imaging.

Major challenges in automated analysis of medical images using deep learning include lack of annotated data and variable presence of anomalies. Manual annotation of tomographic images from techniques such as computed tomography (CT) and magnetic resonance (MR) imaging is typically expertise- and time-intensive. This makes efficient automated volumetric image processing particularly desirable in medical imaging but also presents a major challenge to deep learning research in medical imaging because there are often not enough annotated data for training a deep learning model. In addition, incidental anomalies (e.g. lesions, anatomical variations, imaging artefacts, etc.) apart from any ``primary" pathoanatomy of interest are frequently present in clinical settings, making automated segmentation of multi-component anatomical structures difficult. A method to identify and quantify ``secondary" anomalies will be required for robust segmentation performance.

This work is an extension to preliminary experiments presented in \citet{woo2022anomalyaware} involving segmentation of the distal femur and proximal tibia using U-Net-based networks. In the current work, we develop an approach based on CNN to detect bone anomalies (bone marrow lesions, bone cysts) in the distal femur, proximal tibia and patella from knee MR images from individuals with varying levels of osteoarthritis. The rational was that the output from the anomaly detector can also be utilized for improving segmentation of the knee with osteoarthritis containing bone lesions. The main contributions of this work include:
\begin{enumerate}
\item The use of 3D U-Net-based CNNs for anomaly detection via inpainting in medical imaging. While there have been previous works that used U-Nets for inpainting and visual anomaly detection \citep{liu2020symmetric,ZAVRTANIK2021107706}, they have focused on natural images or 2D image slices. This work extends such ideas to 3D medical images using fully 3D inpainting.
\item Using information learned from anomaly detection to improve segmentation of the distal femur, proximal tibia and patella from knee MR images using an anomaly-aware CNN approach. The current work shows that the proposed approach is capable of significantly improving the segmentation of femur and tibia on osteoarthritic knee MR images with bone anomalies.
\item Demonstrating that the anomaly-aware approach has an advantage when the size of the training dataset is limited. We added additional labels to the segmentation task, including the patella and bone lesions, which are relatively more difficult to segment due to their smaller volumes and greater variability. The proposed anomaly-aware approach is able to detect and segment these new labels with moderate accuracy despite the limited size of the training dataset.
\item Evaluation of the effect of anomaly-aware approach on segmentation performance using two different categories of CNNs: 3D U-Net and 3D context aggregation network (CAN). The current work shows that the anomaly-aware approach can be readily applied to different segmentation CNNs to improve their segmentation performance as well as transfer learning.
\end{enumerate}

\section{Background}
\label{sec:background}

In recent years, several machine learning algorithms have been proposed for automatic anomaly detection. Unsupervised methods using generative models were shown to be promising \citep{baur2018deep,CHEN2020101713} and are particularly useful in medical imaging for which labeled data can be extremely difficult to obtain. A recent work by \citet{PINAYA2022102475} showed that leading-edge techniques such as transformers can also be utilized for anomaly detection and segmentation. However, application of transformers to 3D images is currently challenging due to their very high demands on data and computational resources. Convolutional autoencoders such as U-Net are computationally less intensive than transformers. \citet{PINAYA2022102475} also used vector quantized variational autoencoder (VQ-VAE) for dimensionality reduction which allowed them to apply transformers.

U-Net, developed by \citet{RFB15a}, is a popular CNN model for semantic segmentation. It is a fully convolutional network with an autoencoder structure, consisting of a contracting path (encoder) and an expansive path (decoder). The distinguishing feature of U-Net is the skip connections that transfer feature maps at each resolution from the encoder to the decoder. The skip connections recover spatial information lost during downsampling, which is critical for segmentation tasks. The original U-Net proposed was a 2D convolutional network for segmentation of 2D images. \citet{cciccek20163d} extended this work to dense volumetric segmentation using 3D U-Net. V-Net \citep{milletari2016vnet} is another autoencoder-based model similar to U-Net that is trained end-to-end on image volumes.

Some later works on volumetric segmentation using 3D CNNs (typically U-Net-like) have employed a technique referred to as ``deep supervision" \citep{kayalibay2017cnn,isensee2017brain,raj2018automatic}. The main idea of deep supervision is to provide integrated direct supervision to the hidden layers, rather than providing supervision only at the output layer \citep{pmlr-v38-lee15a}. Deep supervision has been found to speed up convergence likely because it encourages deeper layers to produce improved segmentation results \citep{kayalibay2017cnn}. The skip connections in U-Net force aggregation only at the same-scale feature maps, and it was shown that redesigned skip connections and deep supervision in U-Net enable a significantly higher level of segmentation performance \citep{zhou2020unet++}.

In addition to segmentation, CNNs with U-Net architecture have been applied for different tasks, such as image-to-image translation \citep{isola2017image}. More recently, \citet{liu2020symmetric} adopted a U-Net-based model to perform image inpainting and generate synthetic brain tissue intensities for a tumor region. \citet{ZAVRTANIK2021107706} proposed an anomaly detection method whereby a U-Net-based model was used to reconstruct images from partial inpaintings which were then used to localize visual anomalies in the images. The current work is similar to this idea and perform unsupervised anomaly detection on unlabeled 3D knee MR images through inpainting and lossy reconstruction. The approach is to ``erase" a region of interest (ROI) from an image, which may or may not contain anomalies, and then let the network generate synthetic tissue intensities in the region without the potential anomalies. The difference between the original image and the reconstructed image can then be used to detect the anomalies. Here, we use a version of 3D U-Net for efficient volumetric image processing.

\begin{figure*}[!t]
\centering
\includegraphics[scale=.5]{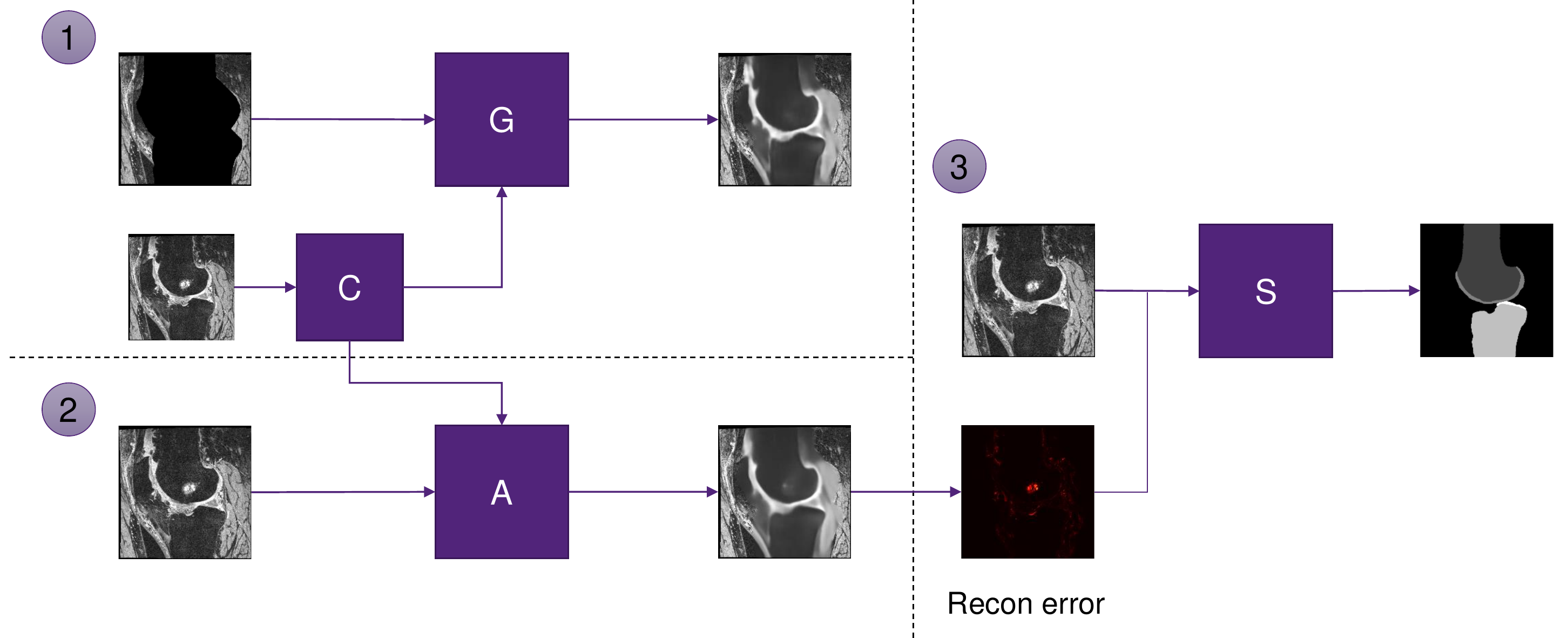}
\caption{Overall pipeline for our method including anomaly detection and the downstream segmentation task. Components \textbf{1} and \textbf{2} are the anomaly detection models (Sections \ref{sec:methods-G} and \ref{sec:methods-A}). Component \textbf{3} is the downstream segmentation task using the anomaly-aware segmentation approach (Section \ref{sec:methods-S}).}
\label{fig:pipeline}
\end{figure*}

As an alternative to autoencoder-based CNNs, context aggregation network (CAN) was proposed by \citet{DBLP:journals/corr/YuK15} for semantic segmentation. It is structurally different from U-Net in that it is not based on an encoder-decoder architecture. In U-Net, the progressive downsampling achieves the effect of integrating contextual information at multi-scale, and the lost resolution is recovered through upsampling. However, since semantic segmentation task requires full-resolution output, there is the question of whether such downsampling and upsampling are truly necessary, which is why \citet{DBLP:journals/corr/YuK15} proposed using dilated convolutions rather than downsampling for multi-scale context aggregation. As with U-Net, CAN was proposed as a 2D network, but \citet{DAI2022102562} extended it to a 3D version for volumetric segmentation.

Accurate segmentation of knee bones and cartilages is an important step in processing and analyzing knee MR images in the context of osteoarthritis. The use of CNNs has grown in knee tissue segmentation to automatically learn image features \citep{EBRAHIMKHANI2020101851}. The first published work in this area is \citet{prasoon2013deep}, which used a triplanar patch-based approach where three CNNs were trained to classify the central voxel as cartilage or background. Although this triplanar approach has now become obsolete, the finding that it achieved a better performance than the previous state-of-the-art showed that a CNN is capable of learning image features from knee MR images.

In our preliminary experiments using 3D U-Net and 3D CAN for knee MR image segmentation, the CNNs were able to achieve mean Dice similarity coefficients (DSCs) for bones and cartilages in the knee joint similar to \citet{AMBELLAN2019109} for the \textit{OAI ZIB} dataset (dataset described in Section \ref{sec:datasets}). However, it was observed that the CNN-based segmentation of images was difficult for cases with visible coexisting abnormalities. Without shape regularization, the surface distance errors tended to be high because CNN outputs often contained ``holes” and ``noises” (false negatives/positives) due to the localized nature of the CNN-based classification. Indeed, \citet{AMBELLAN2019109} used a combination of U-Nets and statistical shape models (SSMs), and the authors explicitly state that SSM regularization of the CNN outputs was needed to attain ``anatomically plausible” segmentations, which is consistent with our preliminary finding. In particular, it was difficult to achieve a good segmentation with a U-Net or CAN when there was a large anomaly in the image (see Figure \ref{fig:segres}). Therefore, we also propose an anomaly-aware segmentation mechanism for CNNs which is more robust in the presence of anomalies.

\section{Methods}
\label{sec:methods}

The overall pipeline for the current work has 3 major components (Figure \ref{fig:pipeline}). In Component 1, the anatomical regions of interest---here, the distal femur and proximal tibia profiles---were erased from the images using the reference segmentation masks, and then these regions were inpainted using a 3D U-Net-based model ($G$). In Component 2, the outputs from Component 1 and another 3D U-Net-based model ($A$) were used to change anomalous bone regions in the original images to close to normal appearances. In Component 3, a 3D CNN-based segmentation network ($S$), which utilizes the information extracted from Component 2, was used to guide the automated segmentation of bone and cartilage volumes compared to the vanilla segmentation networks, specifically aiming to improve segmentation of the structures from images containing visible bone anomalies.

\subsection{Anomaly detection using masked images}
\label{sec:methods-G}

Using the reference segmentation masks available in the \textit{OAI ZIB} MR image dataset (see Section \ref{sec:datasets}), the profiles of the femur and tibia were erased from the MR images. To ensure that the boundaries are not missed and to include some periarticular areas, the masks were dilated by 50 pixels in all directions before being applied to the images. The masked images were then used as input to the 3D U-Net-based model referred to as $G$, which was trained to reconstruct the original image, i.e. to inpaint the erased area.

\begin{figure*}[!t]
\centering
(a)\includegraphics[scale=.5]{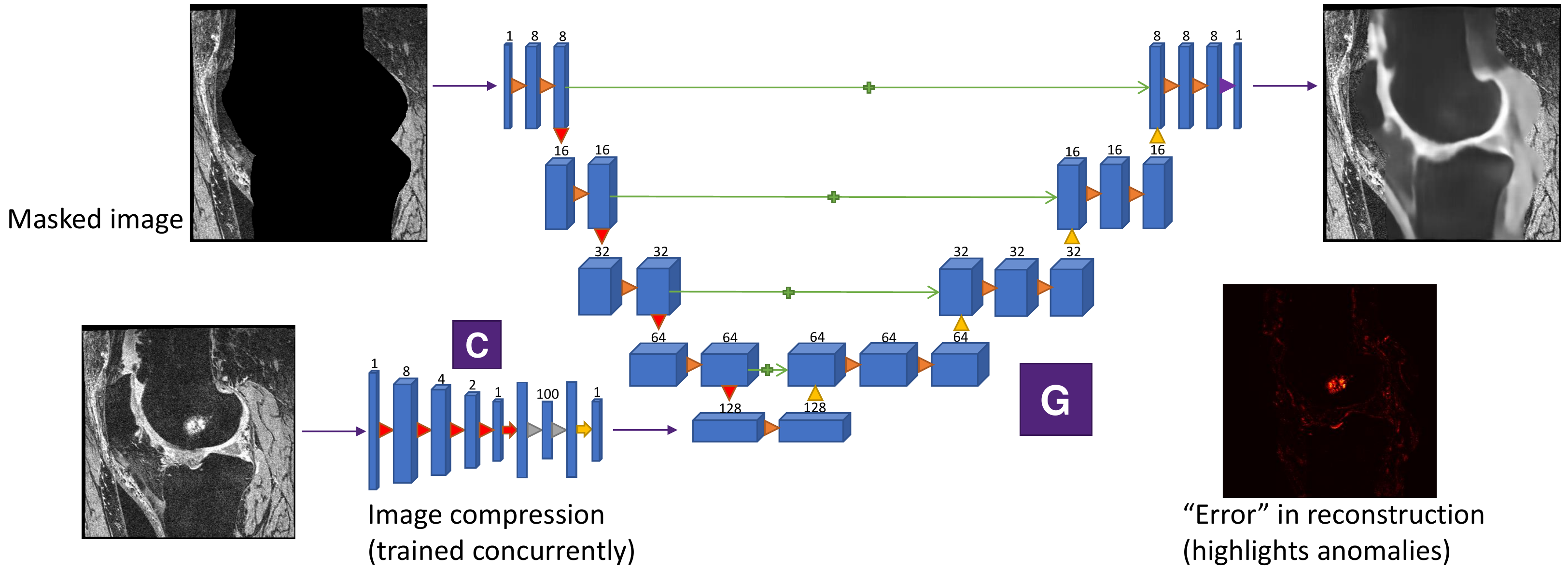}\\
(b)\includegraphics[scale=.5]{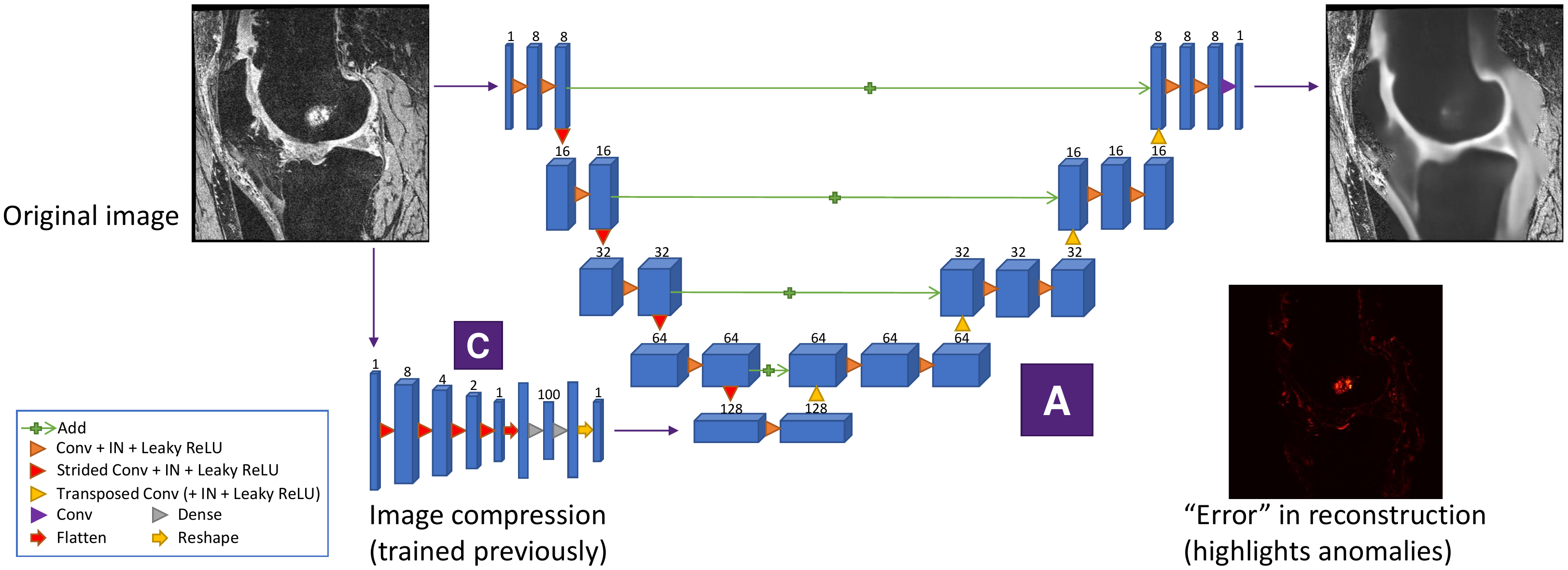}\\
\caption{The two anomaly detection networks with a 3D U-Net-based architecture. Blue boxes represent feature maps, with the number of channels denoted above each box. (a) Network $G$ regenerates the original images from masked images through inpainting and decoding of compressed images. The compressed images are provided by a small network $C$ trained concurrently with $G$. (b) Network $A$ is trained using the images generated by $G$ as the target output given the original images. The compressed images from $C$ are added to the decoder. \textit{IN: Instance Normalization.}}
\label{fig:networks}
\end{figure*}

Since the erased profile was relatively large, an image compressor $C$ was added here to assist with the inpainting. This $C$ was trained simultaneously with $G$ to compress the original MR image, and the compressed image was then fed into the decoder part of $G$. The compressor $C$ had to be a very small network because the bone anomalies must not be recovered in the output; a larger network is likely to restore the anomalies in the inpainted images. Figure \ref{fig:networks}(a) shows the overall structure of the model.

The loss function for training $G$ and $C$ was the mean squared error (MSE) between the original image $x$ and the regenerated image $G(x)$:
\begin{equation}\label{eq:loss-G}
\mathcal{L}_G=||x-G(x)||_2^2.
\end{equation}
It was expected that the model $G$ would recover most of the image but not the anomalous bone regions, and therefore, the squared differences between the original image and the inpainted image $\mathrm{E}(G(x))=(x-G(x))^2$ were used to highlight the anomalies. The region with a larger difference is more likely to be an anomalous region.

\subsection{Anomaly detection using the original images}
\label{sec:methods-A}

A main limitation of the above model $G$ is that it requires a segmentation mask to start with, in order to generate input for the network. Therefore, another network $A$ was trained, which takes the original images (without any masking) as its input and the output images of $G$ as its target output. The network architecture of $A$ was the same as $G$, and the image compressor $C$, trained previously with $G$, was added here as well. Figure \ref{fig:networks}(b) shows the overall structure of the model.

The loss function for training $A$ was the mean squared error (MSE) between the output from the previous network $G(x)$ and the output from the current network $A(x)$, plus the MSE between the respective error images $\mathrm{E}(G(x))=(x-G(x))^2$ and $\mathrm{E}(A(x))=(x-A(x))^2$ to further guide the model:
\begin{equation}\label{eq:loss-A}
\mathcal{L}_A=||G(x)-A(x)||_2^2+||\mathrm{E}(G(x))-\mathrm{E}(A(x))||_2^2.
\end{equation}
Since the outputs of $G$ were used as the target output, it was expected that the model $A$ would produce outputs that are very similar to the outputs from $G$, but it has the main advantage of not requiring segmentation masks to generate the input. This model can be used to detect bone anomalies when given the original images only. Again, the squared differences between the original image and the generated image $\mathrm{E}(A(x))=(x-A(x))^2$ were used to highlight the bone anomalies.

\subsection{Downstream task: Anomaly-aware segmentation}
\label{sec:methods-S}

\begin{figure*}[!t]
\centering
(a)\includegraphics[scale=.5]{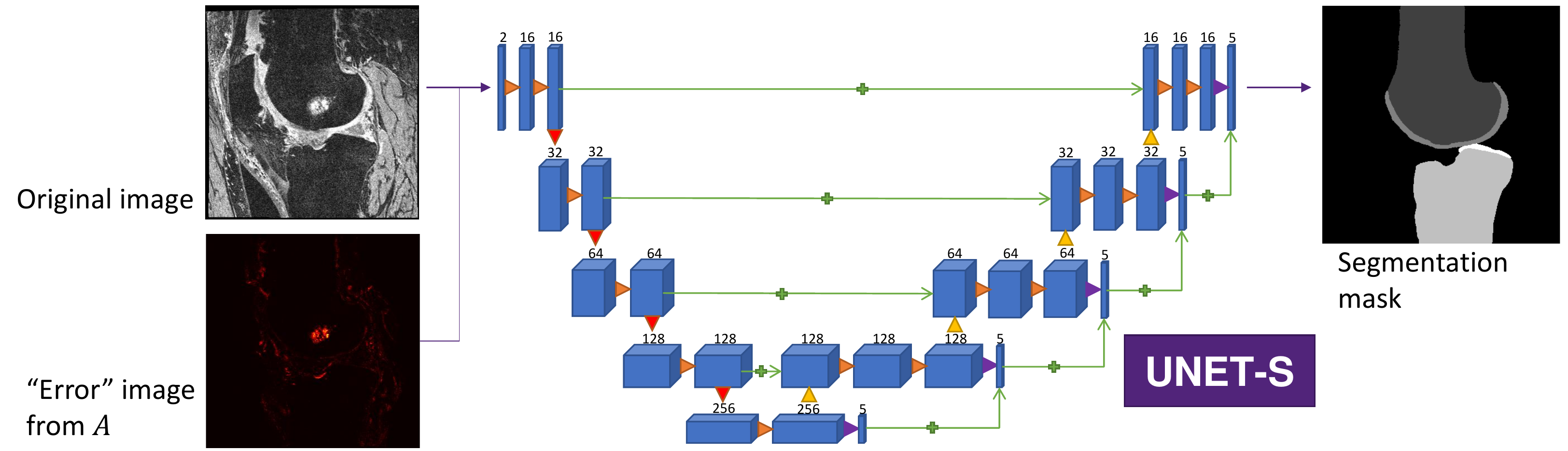}
(b)\includegraphics[scale=.5]{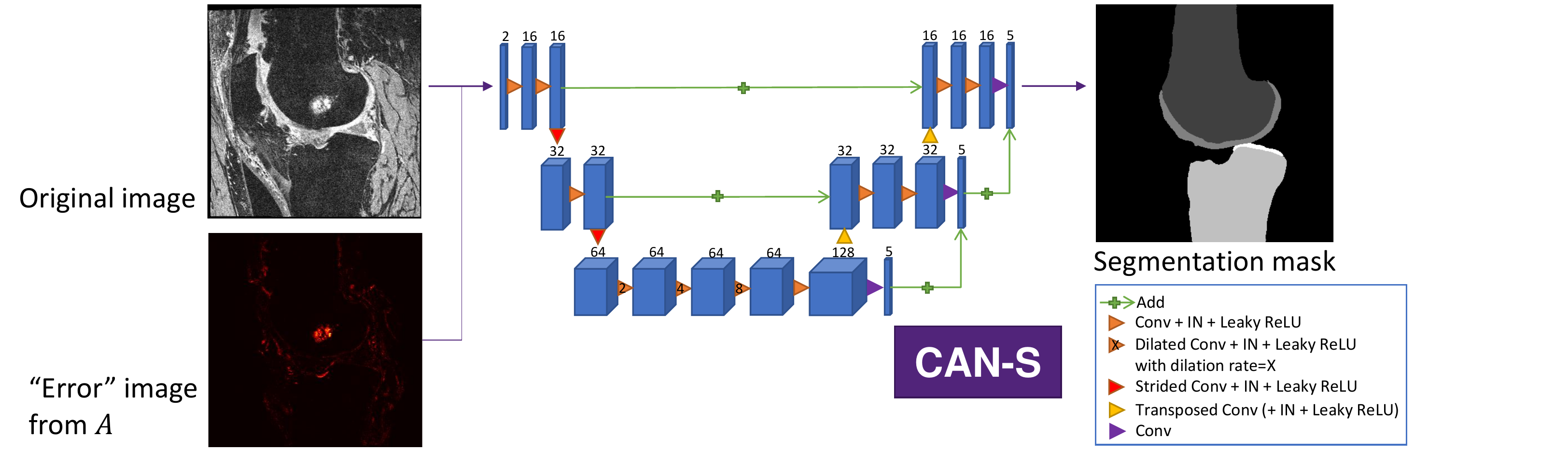}
\caption{The anomaly-aware segmentation network $S$ based on (a) 3D U-Net and (b) 3D CAN with deep supervision. The information extracted from the anomaly detector $A$ was utilized to inform the segmentation of the distal femur and proximal tibia from the knee MR images containing bone abnormalities.}
\label{fig:segnet}
\end{figure*}

To demonstrate the utility of the model $A$, the error images $\mathrm{E}(A(x))$ were utilized to construct segmentation models which can manage anomalies. The anomaly-aware mechanism is a generalized method that can be applied to a CNN-based segmentation network. In this study, the proposed method was tested on two different types of CNNs: 3D U-Net \citep{cciccek20163d} and 3D CAN \citep{DAI2022102562}. Figure \ref{fig:segnet}(a) shows the 3D U-Net architecture, with 4 downsamplings and upsamplings. Figure \ref{fig:segnet}(b) shows the modified 3D CAN. Ideally, CAN would have no downsampling, but it was impossible to use a reasonable number of filters (at least 32) in the full-resolution (160 slices $\times$ 384 $\times$ 384) layers due to the memory limitation of the graphics card, so two downsampling (and upsampling) blocks were added as a trade-off; the CAN module is applied after the two downsamplings. To help stabilize convergence, both networks were modified with deep supervision (see Section \ref{sec:background}) by producing secondary segmentation maps at deeper levels of the network and combining them with the final segmentation map via upsampling and element-wise summation.

Although the original U-Net \citep{RFB15a} used the conventional categorical cross-entropy as the loss function, a multi-class Dice loss is now often used in medical image segmentation because it intrinsically addresses the class imbalance problem commonly seen with medical images \citep{isensee2017brain}. The multi-class Dice loss function is defined as:
\begin{equation}\label{eq:loss-dsc}
\mathcal{L}_{DSC}=1.0-\frac{2}{|K|}\sum_{k \in K}{\frac{\sum_i{u_{i,k}v_{i,k}}}{\sum_i{u_{i,k}}+\sum_i{v_{i,k}}}}.
\end{equation}
Here, $u$ is the softmax output of the network and $v$ is the one-hot encoded ground truth segmentation map; $K$ is the number of classes, and $u_{i,k}$ and $v_{i,k}$ denote the softmax output and ground truth label, respectively, for class $k$ at voxel $i$.

While the conventional U-Net is able to produce plausible segmentation results for most images without anomalies, it can easily fail when there are anomalies in the images since anomalies were not taken into account. To address this problem, the error images $\mathrm{E}(A(x))$ were added as an additional input (Figure \ref{fig:segnet}) along with an additional loss function. Inspired by the work of \citet{nie2020adversarial}, which used a difficulty-aware attention mechanism, a focal cross-entropy loss was added to the loss function, where the focal weights were given by the error images:
\begin{equation}\label{eq:loss-fce}
\mathcal{L}_{FCE}=-\sum_i{\sum_{k \in K}{F_iv_{i,k}\log{u_{i,k}}}},
\end{equation}
where $F=1.0+\beta\mathrm{E}(A(x))$ and $F_i$ denotes the focal weight at voxel $i$. Here, a weighting factor of $\beta=99.0$ was used so that the values of $F$ ranges from 1 to 100. (If $\beta=0.0$, Equation \ref{eq:loss-fce} would be the same as the usual categorical cross-entropy.)

The total loss for the segmentation network was then:
\begin{equation}\label{eq:loss-S}
\mathcal{L}_S=\mathcal{L}_{DSC}+\alpha\mathcal{L}_{FCE},
\end{equation}
where $\alpha$ is another weighting factor. Here, $\alpha=10.0$ was used. Using this loss, the segmentation network can be trained to pay more attention to the bone voxels that were found to be anomalous by the network $A$ and hence likely to be difficult for the segmentation network to classify. For a segmentation task, it is assumed that segmentation masks are not available for test images, so only the outputs from $A$ (not $G$), which did not require segmentation masks, would be used for training and testing the segmentation network.

\begin{table*}[!t]
\caption{Summary of the datasets used in the current study.}

\begin{center}
\begin{tabular}{|c|c|c|c|}
\hline
      ~& \textbf{OAI ZIB}& \textbf{OAI ZIB--UQ}& \textbf{OAI AKOA}\\
\hline
      Number of subjects& 507& 20& 24\\
\hline
      Timepoints& baseline& baseline& baseline \& last follow-up\\
\hline
      Manual segmentations& FB, FC, TB, TC& \multicolumn{2}{|c|}{FB, FC, TB, TC, PB, PC, FL, TL PL}\\
\hline
      \multirow{2}{*}{Used for}& training $G$& \multirow{2}{*}{training $S^T$}& testing $G$ \& $A$\\
      ~& training \& testing (CV) $A$ and $S$& ~& training \& testing (CV) $S^T$\\
\hline
\multicolumn{4}{@{}l}{\textit{FB: femoral bone; FC: femoral cartilage; TB: tibial bone; TC: tibial cartilage; PB: patellar bone; PC:}}\\
\multicolumn{4}{@{}l}{\textit{patellar cartilage; FL: femoral lesion; TL: tibial lesion; PL: patellar lesion; CV: cross-validation}}
\end{tabular}
\end{center}

\label{tab:datasets}
\end{table*}

In this paper, we will refer to our 3D U-Net and 3D CAN modified with deep supervision and anomaly-aware mechanism as $UNET$-$S$ and $CAN$-$S$, respectively.

\section{Experiments}
\label{sec:experiments}

We trained and tested our method on subsets of knee MR images from the publicly available Osteoarthritis Initiative (OAI) database \citep{OAI}. Three different subsets were used: \textit{OAI ZIB}, \textit{OAI ZIB--UQ}, and \textit{OAI AKOA} (summarized in Table \ref{tab:datasets}).

\subsection{MR image datasets}
\label{sec:datasets}

This study used the publicly available knee MR image dataset \textit{OAI ZIB}, generated by researchers at Zuse Institute Berlin (ZIB) \citep{AMBELLAN2019109}. The dataset consists of 507 MR examinations from the OAI for which manual reference segmentations of femoral and tibial bones and cartilages were produced by experienced analysts starting from a model-based auto-segmentation. The images are 160 sagittal slices $\times$ 384 $\times$ 384 voxels, and they are all images of the right knee at baseline. The MR imaging sequence is 3D DESS (double-echo steady state) with water excitation. Segmentation labels consist of the background (0), femoral bone (1), femoral cartilage (2), tibial bone (3), and tibial cartilage (4). (Background refers to all voxels that were not labeled specifically.) The dataset covers the full spectrum of osteoarthritis grades. Further details of the dataset can be found in \citet{AMBELLAN2019109}.

The public \textit{OAI ZIB} dataset is relatively large, but it has segmentation labels for femoral and tibial bones and cartilages only. As a pilot study, we added some additional segmentation labels to 20 MR examinations from the \textit{OAI ZIB} dataset with varying osteoarthritis severity. The additional segmentation labels included patellar bone (5) and patellar cartilage (6). In addition, bone marrow lesions (BMLs) or subchondral cysts for each bone---i.e. femoral lesion (7), tibial lesion (8), and patellar lesion (9)---were also segmented. We will refer to these 20 MR examinations as the \textit{OAI ZIB--UQ} dataset.

A new dataset called \textit{OAI AKOA} with 9 segmentation labels (10 including image background) was also produced. The dataset contains images from the OAI database with ``accelerated" knee osteoarthritis, defined as patients who had Kellgren–Lawrence (KL) grade $\leq$ 1 at baseline but progressed to KL grade $\geq$ 2 within the follow-up period. The \textit{OAI AKOA} dataset consists of 48 MR examinations from 24 patients acquired at 2 timepoints, (baseline and last imaging follow-up at a maximum of 96 months). The segmentations for the \textit{OAI ZIB--UQ} and \textit{OAI AKOA} datasets were carried out manually using ITK-SNAP \citep{py06nimg} by an analyst (WB) with supervision from another analyst (CE) with expertise in segmentation of the human musculoskeletal system.

The manual segmentations in the \textit{OAI ZIB--UQ} and \textit{OAI AKOA} datasets as well as the cross-validation data splits (\textit{vide infra}) are released in \url{https://github.com/wooboyeong/Anomaly-Aware-3D-Segmentation}. Ethics approval for collection of human data was obtained by the OAI and the participating clinical sites \citep{OAI}.

\subsection{Experimental setup}
\label{sec:setup}

\begin{figure}[!b]
\centering
\includegraphics[scale=.45]{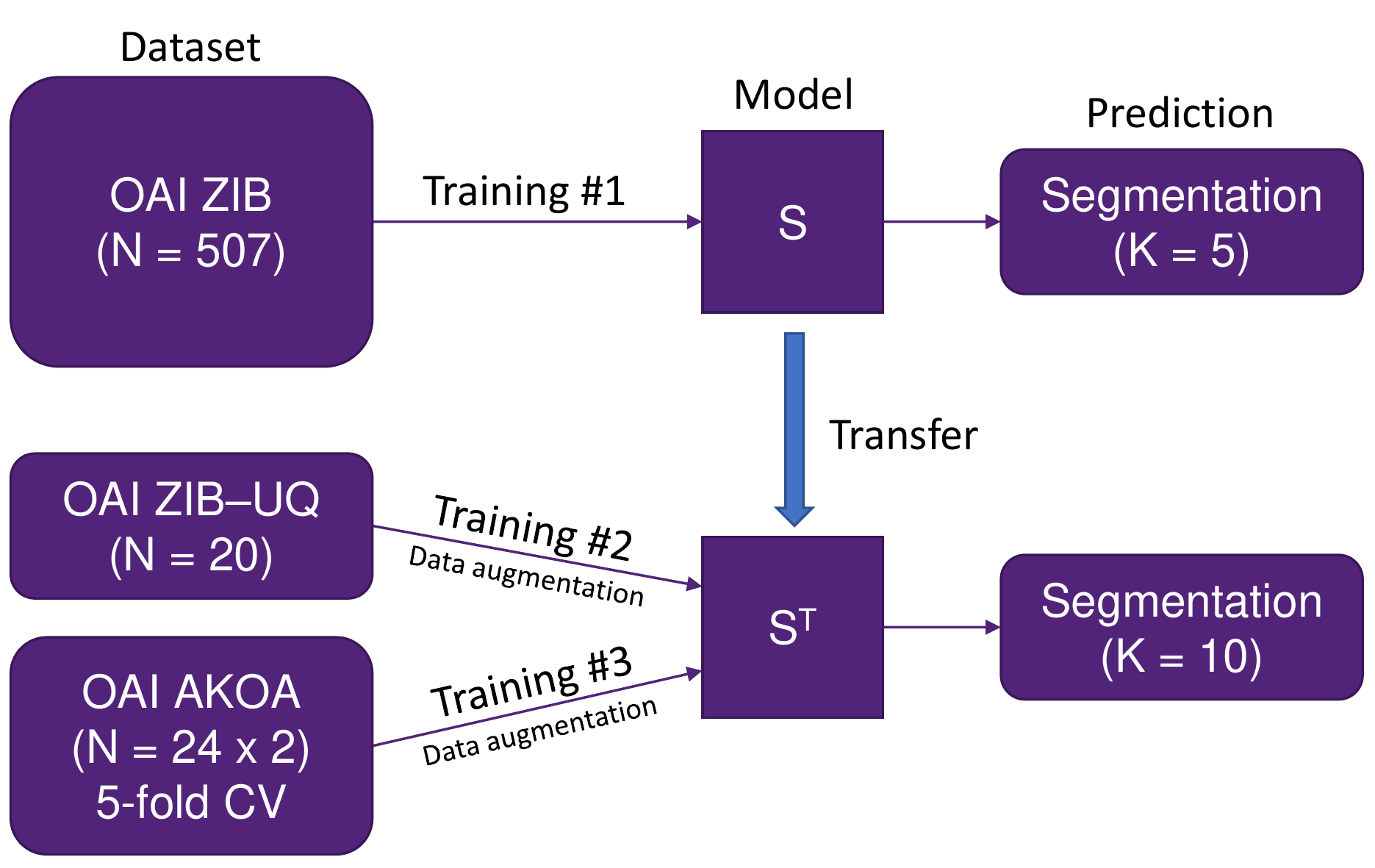}
\caption{Transfer learning for further segmentation of knee MR images. Here, $N$ refers to the number of MR examinations while $K$ refers to the number of segmentation classes (including the background) in the dataset. See Table \ref{tab:datasets} for the list of segmentation classes.}
\label{fig:transfer}
\end{figure}

\begin{figure*}[!t]
\centering
\includegraphics[scale=.6]{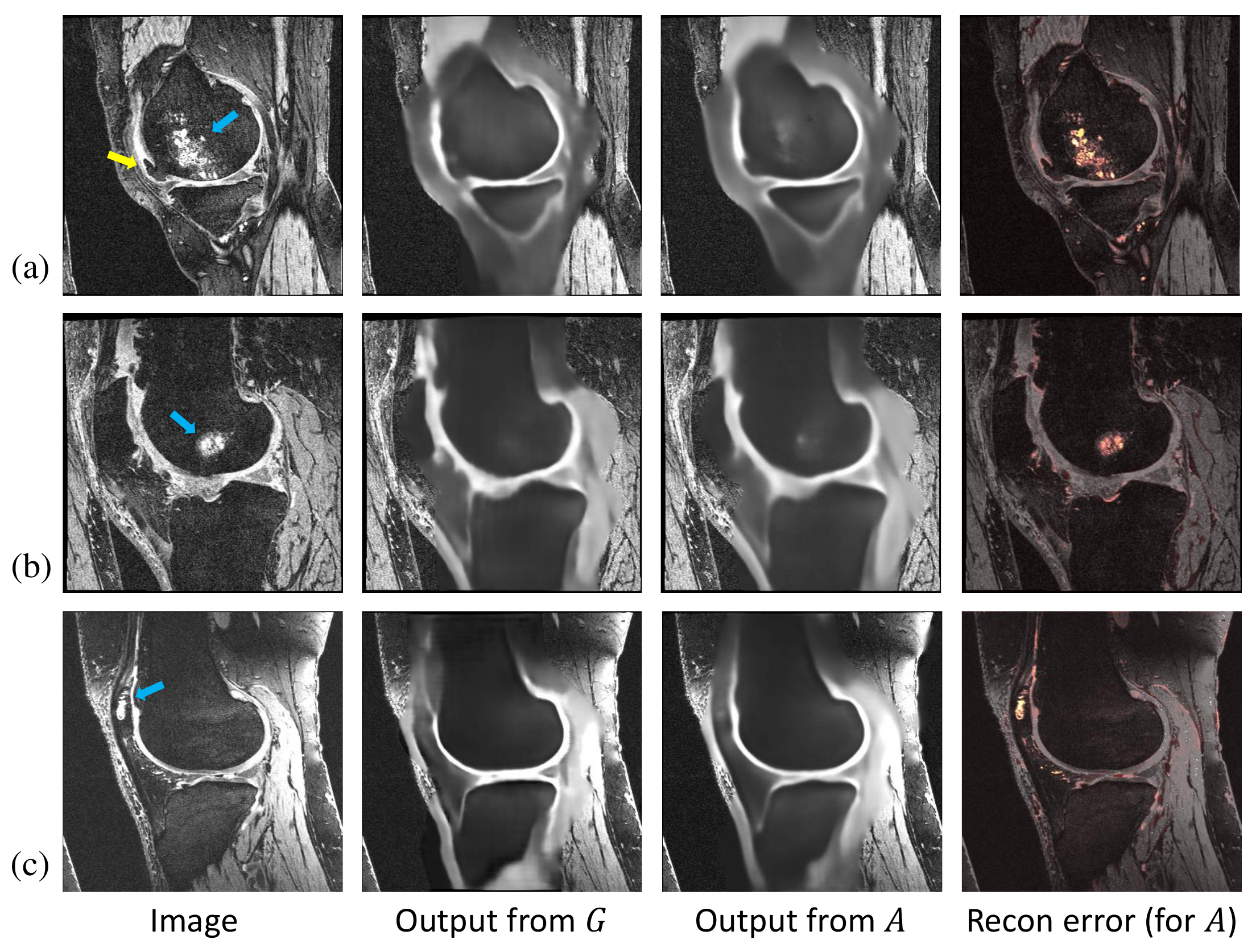}
\caption{Example outputs from the bone anomaly detection networks $G$ and $A$. Figures (a) and (b) are images from the \textit{OAI ZIB} dataset, and Figure (c) is an image from the \textit{OAI AKOA} dataset. The last column shows the error images (color-mapped and overlaid on the input images) highlighting the difference between the input image and the output from $A$. Regions of BMLs (\textit{blue arrows}) in the femur and patella and part of an osteophyte (\textit{yellow arrow}) on the femur had high reconstruction errors.}
\label{fig:recon}
\end{figure*}

\begin{figure*}[!t]
\centering
\includegraphics[scale=.54]{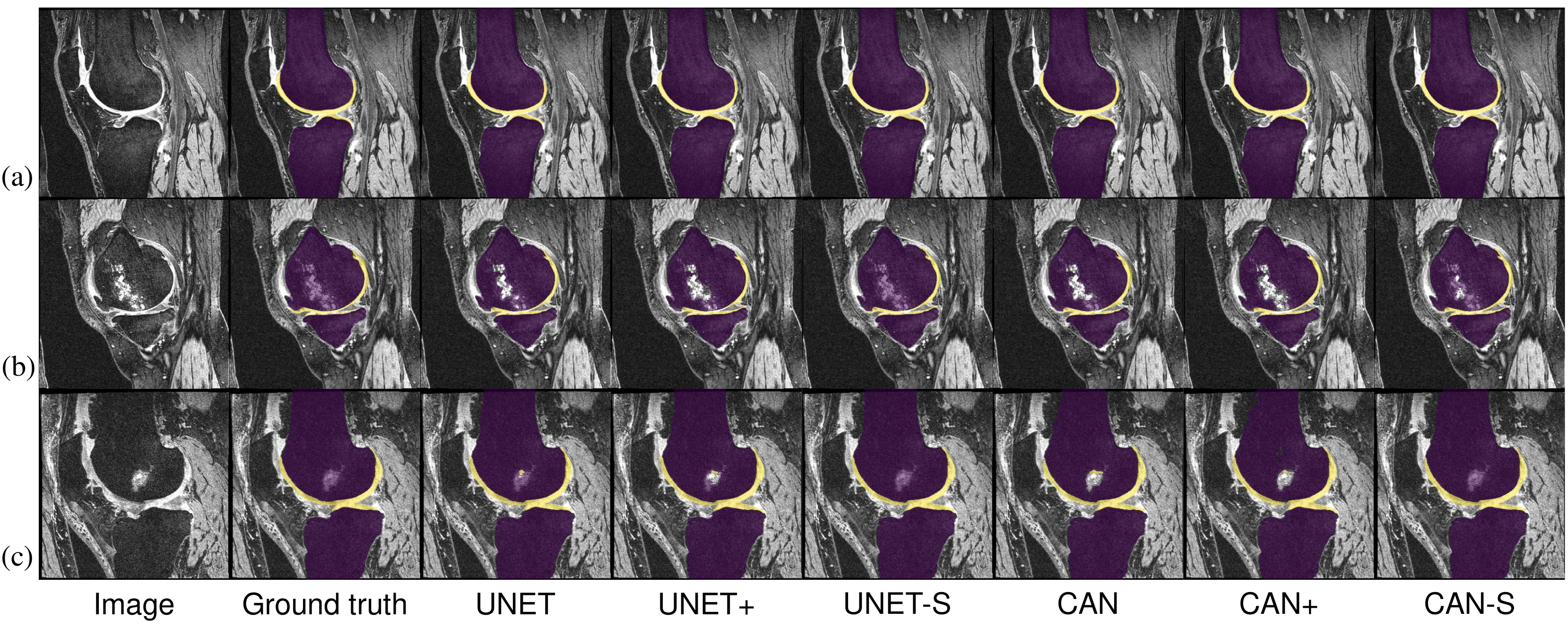}
\caption{Example segmentation outputs for the femoral and tibial bones (\textit{purple}) and cartilages (\textit{yellow}) generated by the individual network implementation with the \textit{OAI ZIB} dataset. The examples show (a) images with little to no visible bone anomalies where all networks produced good segmentation masks and (b,c) images with visible bone anomalies where segmentation networks tended to fail to produce plausible segmentation masks. The anomaly-aware networks, especially $UNET$-$S$, were better able to correctly segment the images with anomalies.}
\label{fig:segres}
\end{figure*}

To evaluate the anomaly-aware segmentation method, the segmentation networks $UNET$-$S$ and $CAN$-$S$ were compared with the standard 3D U-Net and 3D CAN without any modification, as well as the models with deep supervision only (without the anomaly-aware mechanism). The loss function for these networks was multi-class Dice loss only and there was no additional input. The network architecture and the training setup were otherwise the same as $UNET$-$S$ and $CAN$-$S$. For convenience, the models without deep supervision will be referred to as $UNET$ and $CAN$, and those with deep supervision will be referred to as $UNET+$ and $CAN+$.  See \ref{sec:implementation} for implementation and training details.

\begin{table*}[!t]
\caption{Mean DSC, ASD, and HD values for segmentations of the femoral and tibial bone and cartilage volumes from the proposed anomaly-aware method ($UNET$-$S$ and $CAN$-$S$) with their baseline networks, evaluated using 5-fold cross-validation on the \textit{OAI ZIB} dataset ($N$ = 507).}

\begin{center}
\begin{tabular}{|c|c||c|c|c||c|c|c|}
\hline
      \textbf{Class}& \textbf{Metric}& \textbf{UNET}& \textbf{UNET+}& \textbf{UNET-S}& \textbf{CAN}& \textbf{CAN+}& \textbf{CAN-S}\\
\hline
      ~& DSC (\%)& 98.6$\pm$0.33& 98.7$\pm$0.33& \textbf{\underline{98.7$\pm$0.30}}& 98.6$\pm$0.37$^\dag$& 98.6$\pm$0.35$^\dag$& \textbf{98.7$\pm$0.30}\\
      ~& ASD (mm)& 0.23$\pm$0.06& 0.24$\pm$0.08$^\dag$& \textbf{\underline{0.22$\pm$0.05}}& 0.25$\pm$0.14$^{\dag\ddag}$& 0.25$\pm$0.09$^{\dag\ddag}$& \textbf{0.23$\pm$0.07}\\
      \multirow{2}{*}{\textbf{FB}}& HD (mm)& 7.05$\pm$6.01$^{\dag\ddag}$& 9.96$\pm$7.06$^{\dag\ddag}$& \textbf{\underline{3.40$\pm$2.67}}& 10.07$\pm$7.53$^{\dag\ddag}$& 11.33$\pm$7.67$^{\dag\ddag}$& \textbf{4.22$\pm$3.96}\\
      ~& HD$^*$ (mm)& 55.54$\pm$35.29$^{\dag\ddag}$& 24.18$\pm$19.96$^{\dag\ddag}$& \textbf{\underline{4.05$\pm$5.94}}& 71.78$\pm$26.49$^{\dag\ddag}$& 28.28$\pm$19.94$^{\dag\ddag}$& \textbf{5.82$\pm$8.36}\\
\cline{2-8}
      ~& HD$_0$ (mm)& 5.97$\pm$5.39$^{\dag\ddag}$& 9.46$\pm$7.33$^{\dag\ddag}$& \textbf{\underline{2.82$\pm$1.63}}& 8.98$\pm$7.14$^{\dag\ddag}$& 10.24$\pm$7.76$^{\dag\ddag}$& \textbf{3.46$\pm$2.71}\\
      ~& HD$_1$ (mm)& 7.74$\pm$6.29$^{\dag\ddag}$& 10.29$\pm$6.87$^{\dag\ddag}$& \textbf{\underline{3.78$\pm$3.10}}& 10.76$\pm$7.71$^{\dag\ddag}$& 12.03$\pm$7.54$^{\dag\ddag}$& \textbf{4.71$\pm$4.52}\\
\hline
      ~& DSC (\%)& \textbf{89.6$\pm$2.79$^\ddag$}& \textbf{\underline{89.7$\pm$2.91$^\ddag$}}& 89.5$\pm$2.67& 89.2$\pm$2.71& 89.3$\pm$2.69& 89.0$\pm$2.46\\
      ~& ASD (mm)& 0.26$\pm$0.07& \textbf{\underline{0.26$\pm$0.07}$^\ddag$}& \textbf{0.26$\pm$0.07}& 0.27$\pm$0.07$^\dag$& 0.27$\pm$0.07& 0.27$\pm$0.07\\
      \multirow{2}{*}{\textbf{FC}}& HD (mm)& 7.44$\pm$4.49$^{\dag\ddag}$& 6.71$\pm$3.99$^{\dag\ddag}$& \textbf{\underline{5.29$\pm$2.39}}& 8.46$\pm$4.89$^{\dag\ddag}$& 7.17$\pm$4.15$^{\dag\ddag}$& \textbf{5.57$\pm$2.63}\\
      ~& HD$^*$ (mm)& 14.38$\pm$14.85$^{\dag\ddag}$& 14.06$\pm$15.04$^{\dag\ddag}$& \textbf{\underline{5.58$\pm$3.69}}& 34.82$\pm$30.24$^{\dag\ddag}$& 16.74$\pm$16.18$^{\dag\ddag}$& \textbf{6.05$\pm$4.31}\\
\cline{2-8}
      ~& HD$_0$ (mm)& 6.51$\pm$4.25$^{\dag\ddag}$& 5.59$\pm$3.06$^\dag$& \textbf{\underline{4.52$\pm$1.60}}& 7.48$\pm$4.59$^{\dag\ddag}$& 5.92$\pm$3.51$^{\dag\ddag}$& \textbf{4.85$\pm$1.95}\\
      ~& HD$_1$ (mm)& 8.04$\pm$4.54$^{\dag\ddag}$& 7.43$\pm$4.35$^{\dag\ddag}$& \textbf{\underline{5.78$\pm$2.68}}& 9.09$\pm$4.98$^{\dag\ddag}$& 7.98$\pm$4.33$^{\dag\ddag}$& \textbf{6.03$\pm$2.89}\\
\hline
      ~& DSC (\%)& 98.7$\pm$0.35& \textbf{98.7$\pm$0.33}& \textbf{\underline{98.7$\pm$0.32}}& 98.6$\pm$0.35$^\dag$& 98.6$\pm$0.36$^\dag$& 98.7$\pm$0.33\\
      ~& ASD (mm)& 0.22$\pm$0.10& \textbf{0.22$\pm$0.06}& \textbf{\underline{0.21$\pm$0.06}}& 0.23$\pm$0.08$^\dag$& 0.24$\pm$0.10$^{\dag\ddag}$& 0.22$\pm$0.08\\
      \multirow{2}{*}{\textbf{TB}}& HD (mm)& 6.29$\pm$5.72$^{\dag\ddag}$& 7.03$\pm$5.23$^{\dag\ddag}$& \textbf{\underline{3.23$\pm$1.86}}& 9.26$\pm$7.03$^{\dag\ddag}$& 10.09$\pm$7.23$^{\dag\ddag}$& \textbf{3.85$\pm$3.16}\\
      ~& HD$^*$ (mm)& 76.04$\pm$45.14$^{\dag\ddag}$& 27.63$\pm$26.23$^{\dag\ddag}$& \textbf{\underline{3.82$\pm$5.63}}& 65.28$\pm$37.62$^{\dag\ddag}$& 37.06$\pm$24.76$^{\dag\ddag}$& \textbf{5.90$\pm$10.58}\\
\cline{2-8}
      ~& HD$_0$ (mm)& 5.59$\pm$4.97$^{\dag\ddag}$& 5.85$\pm$4.37$^{\dag\ddag}$& \textbf{\underline{3.01$\pm$1.82}}& 8.44$\pm$6.81$^{\dag\ddag}$& 9.94$\pm$7.64$^{\dag\ddag}$& \textbf{3.12$\pm$1.94}\\
      ~& HD$_1$ (mm)& 6.73$\pm$6.11$^{\dag\ddag}$& 7.78$\pm$5.59$^{\dag\ddag}$& \textbf{\underline{3.38$\pm$1.87}}& 9.78$\pm$7.13$^{\dag\ddag}$& 10.19$\pm$6.96$^{\dag\ddag}$& \textbf{4.32$\pm$3.67}\\
\hline
      ~& DSC (\%)& \textbf{85.9$\pm$4.21}& 85.8$\pm$4.16& \textbf{\underline{86.0$\pm$4.00}}& 85.3$\pm$4.11& 85.2$\pm$4.24$^\dag$& 85.3$\pm$4.17\\
      ~& ASD (mm)& 0.27$\pm$0.10& \textbf{0.27$\pm$0.10}& \textbf{\underline{0.26$\pm$0.09}}& 0.28$\pm$0.10$^\dag$& 0.28$\pm$0.10& 0.27$\pm$0.11\\
      \multirow{2}{*}{\textbf{TC}}& HD (mm)& 5.83$\pm$3.23$^{\dag\ddag}$& 5.43$\pm$2.83$^\dag$& \textbf{\underline{4.74$\pm$2.03}}& 7.00$\pm$4.15$^{\dag\ddag}$& 6.09$\pm$3.36$^{\dag\ddag}$& \textbf{4.89$\pm$2.20}\\
      ~& HD$^*$ (mm)& 11.47$\pm$16.26$^{\dag\ddag}$& 8.75$\pm$12.13$^{\dag\ddag}$& \textbf{\underline{4.74$\pm$2.03}}& 16.38$\pm$17.96$^{\dag\ddag}$& 14.23$\pm$17.93$^{\dag\ddag}$& \textbf{4.99$\pm$2.81}\\
\cline{2-8}
      ~& HD$_0$ (mm)& 4.70$\pm$2.18& 4.69$\pm$2.29& \textbf{\underline{4.19$\pm$1.40}}& 6.01$\pm$3.42$^{\dag\ddag}$& 5.22$\pm$2.78$^{\dag\ddag}$& \textbf{4.32$\pm$1.52}\\
      ~& HD$_1$ (mm)& 6.56$\pm$3.58$^{\dag\ddag}$& 5.91$\pm$3.04$^\dag$& \textbf{\underline{5.09$\pm$2.27}}& 7.63$\pm$4.45$^{\dag\ddag}$& 6.65$\pm$3.58$^{\dag\ddag}$& \textbf{5.26$\pm$2.48}\\
\hline
\multicolumn{8}{@{}l}{\small Bold with underline represents the best value within each metric. Bold without underline represents the second best value.}\\
\multicolumn{8}{@{}l}{\small HD$^*$ refers to Hausdoff distances before post-processing. All other metrics refer to results after post-processing.}\\
\multicolumn{8}{@{}l}{\small HD$_0$ refers to HDs (after post-processing) for cases with mild to no osteoarthritis (radiographic grade $\le$ 2; $N_0$ = 198).}\\
\multicolumn{8}{@{}l}{\small HD$_1$ refers to HDs (after post-processing) for cases with moderate to severe osteoarthritis (radiographic grade $\ge$ 3; $N_1$ = 309).}\\
\multicolumn{8}{@{}l}{\small $^\dag$ represents significant difference (p-value $<$ 0.05) compared to UNET-S with Tukey's HSD test.}\\
\multicolumn{8}{@{}l}{\small $^\ddag$ represents significant difference (p-value $<$ 0.05) compared to CAN-S with Tukey's HSD test.}
\end{tabular}
\end{center}

\label{tab:segres}
\end{table*}

The \textit{OAI ZIB} dataset was randomly and evenly split into 5-fold cross-validation sets, i.e. 102/102/101/101/101 MR examinations in each test set, and the split was maintained over all parts of Components 2 ($A$) and 3 ($S$) in Figure \ref{fig:pipeline}, as well as over all different segmentation models being evaluated. ($S$ refers to either $UNET$--$S$ or $CAN$--$S$.) Segmentation performance was evaluated using Dice similarity coefficient (DSC), average surface distance (ASD), and Hausdorff distance (HD) values. See \ref{sec:evaluation} for the definitions of these evaluation metrics.

\subsection{Transfer learning for further segmentation}
\label{sec:transfer}

As an extension to the above experiments, the segmentation networks were trained further with the expanded number of segmentation labels in the \textit{OAI ZIB--UQ} and \textit{OAI AKOA} datasets. Since these datasets are relatively small, the segmentation networks were first trained on the entire \textit{OAI ZIB} dataset, and then the learned weights were transferred to new networks (Figure \ref{fig:transfer}). The new networks have 10 channels in the output layer instead of 5 since there are now 10 segmentation classes (including image background). The new networks will be denoted with the superscript $^T$.

The Dice loss function (see Equation \ref{eq:loss-dsc}) was modified so that higher weights were given to the new classes that the network now has to learn:
\begin{equation}\label{eq:loss-wdsc}
\mathcal{L}_{wDSC}=\frac{1}{|K|}\sum_{k \in K}w_k\left(1.0-\frac{2\sum_i{u_{i,k}v_{i,k}}}{\sum_i{u_{i,k}}+\sum_i{v_{i,k}}}\right),
\end{equation}
where $w_k=1.0$ for the first 5 classes that were already learned, and $w_k=10.0$ for the new 5 classes. In addition, since many images do not contain lesions, $w_k$ was changed to 0 if $\sum_i{u_{i,k}}+\sum_i{v_{i,k}}=0.0$; this means the image did not contain label $k$ and the network successfully predicted that there is no label $k$, so the loss in this case would be 0. The loss function for the baseline networks ($UNET^T$, $UNET+^T$, $CAN^T$, $CAN+^T$) was this weighted multi-class Dice loss.

The focal cross-entropy loss for the anomaly-aware networks  (see Equation \ref{eq:loss-fce}) was the same as before. Therefore, the total loss for $UNET$-$S^T$ and $CAN$-$S^T$ was:
\begin{equation}\label{eq:loss-ST}
\mathcal{L}_{S^T}=\mathcal{L}_{wDSC}+\alpha\mathcal{L}_{FCE}.
\end{equation}

The new networks were then trained on the entire \textit{OAI ZIB--UQ} dataset to learn to segment the 10 classes. Lastly, in order to test the pipeline $A \rightarrow S^T$, the anomaly information for images in the \textit{OAI AKOA} dataset was obtained by running the images through the network $A$ that was trained on the \textit{OAI ZIB} dataset. Since $A$ only requires the original images for reconstruction, it did not need any further training or transfer learning. However, since the segmentation networks only had 20 images (the \textit{OAI ZIB--UQ} dataset) for learning the new labels, it required some further training, so the \textit{OAI AKOA} dataset was randomly and evenly split into 5-fold cross-validation sets on a patient basis, i.e. 5$\times$2/5$\times$2/5$\times$2/5$\times$2/4$\times$2 testing images in each set. Due to the limited size of the datasets, online data augmentation was applied to reduce overfitting.

Figure \ref{fig:transfer} shows a summary of the transfer learning process. For all of the networks, weights for the first two convolutional layers were frozen during training (see Figure \ref{fig:segnet-T} in the Appendix). See \ref{sec:implementation} for implementation and training details.

\section{Results}
\label{sec:results}

\subsection{Anomaly detection}
\label{sec:results-detection}

Figure \ref{fig:recon} shows example output images from the anomaly detection networks $G$ and $A$. The input MR images of the knee have some visible bone anomalies including BMLs and osteophytes. The network outputs are lossy reconstructions of the input images with the bright signal bone anomalies within the cancellous bone mostly removed from the images. Some of the osteophytes were incompletely reconstructed. The network $A$ only had the original images as inputs, but the outputs from $A$ still have most of the anomalies blurred out. The last column of Figure \ref{fig:recon} shows the reconstruction error images from $A$ in which the anomalous regions detected within the cancellous bones are highlighted.

Note that the MR image in Figure \ref{fig:recon}(c) is from the \textit{OAI AKOA} dataset, which was not used for training either $G$ or $A$, but only the masks of femur and tibia were used to generate input for $G$ since the model was only trained with the \textit{OAI ZIB} dataset. However, since the masks had been dilated before erasing the images (see Section \ref{sec:methods-G}), the subchondral patellar lesions were also detected, as can be seen in Figure \ref{fig:recon}(c). Also note that masks are unnecessary for $A$ in any case.

\subsection{Anomaly-aware segmentation on OAI ZIB dataset}
\label{sec:results-segmentation}

Figure \ref{fig:segres} shows example outputs from the anomaly-aware segmentation networks $UNET$-$S$ and $CAN$-$S$ and also the outputs from their baseline networks. As noted in Section \ref{sec:background}, the CNNs performed well in terms of DSCs for most of the \textit{OAI ZIB} images (Figure \ref{fig:segres}(a)), but failed on some cases with severe abnormalities. The anomaly-aware method was found to be more robust against these difficult cases, resulting in a noticeable improvement in the quality of the segmentation of bone volume for the femur and tibia (Figures \ref{fig:segres}(b) and \ref{fig:segres}(c)). The image in Figure \ref{fig:segres}(b) was the most difficult from the \textit{OAI ZIB} dataset for the segmentation CNNs due to the presence of a large femoral BML, but the segmentation error was fixed with $UNET$-$S$ and partially fixed with $CAN$-$S$. The image in Figure \ref{fig:segres}(c) also has a notable femoral BML, but both $UNET$-$S$ and $CAN$-$S$ were able to correctly segment the femoral bone.

Table \ref{tab:segres} shows the quantitative results for the segmentation task. Tukey's honestly significant difference (HSD) test was used to compare all possible pairs of means for each metric; this test is is similar to the \textit{t}-test, except it corrects for family-wise error rate. There was not a significant improvement in the mean DSCs, but HDs were substantially reduced for all of the segmentation classes for both $UNET$-$S$ and $CAN$-$S$ compared to their baselines. The HDs for cases with moderate to severe osteoarthritis (HD$_1$ in Table \ref{tab:segres}) were higher than the HDs for cases with mild to no osteoarthritis (HD$_0$) for all models, but the HD$_1$ was still relatively low with the anomaly-aware models compared to the other models.

Note that the DSCs, ASDs, and HDs were calculated after post-processing the CNN outputs. The post-processing was performed because most of the CNN outputs for non-anomaly-aware models contained random stray voxels which resulted in extremely high HDs (see HD$^*$ in Table \ref{tab:segres}). As mentioned in Section \ref{sec:background}, this is one of the main limitations of CNN-based segmentations. The random stray voxels were removed by extracting the largest components and removing smaller objects that are distant from the main structures. See \ref{sec:post-processing} for description of the post-processing method. The mean HDs were significantly reduced after post-processing for the non-anomaly-aware models. The CNN outputs from the anomaly-aware models did not contain as many stray voxels to start with, so the effect of post-processing was much less obvious.

\begin{figure}[!t]
\centering
\includegraphics[scale=.43]{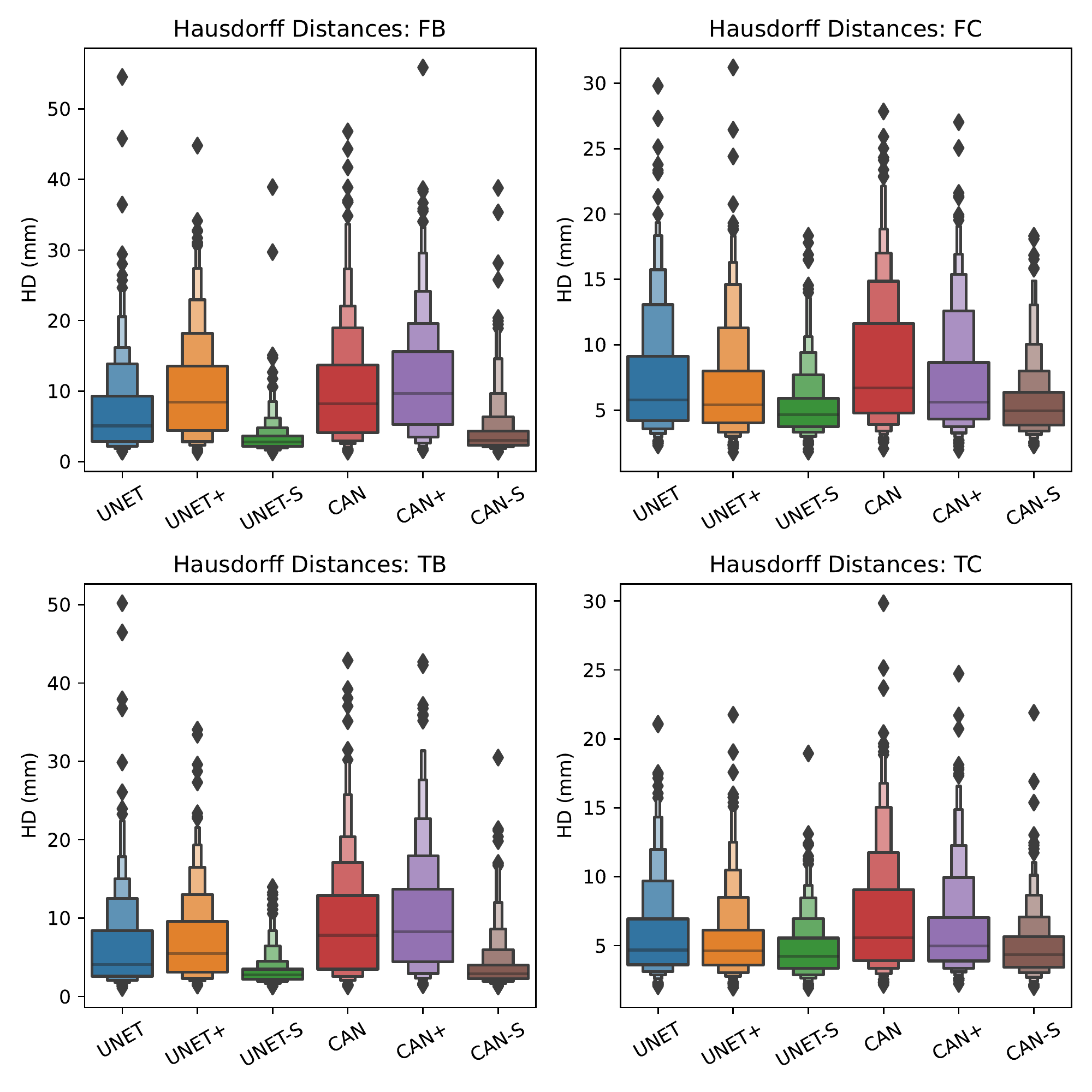}
\caption{Boxplots of Hausdorff distance (HD) values for the proposed anomaly-aware segmentation approach ($UNET$-$S$ and $CAN$-$S$) and baseline networks, evaluated on the \textit{OAI ZIB} dataset using 5-fold cross-validation. Note that these HDs are results after post-processing.}
\label{fig:boxplots}
\end{figure}

Even after the post-processing, the mean HDs for the anomaly-aware models were significantly less than the mean HDs for the non-anomaly-aware models. See also the boxplots in Figure \ref{fig:boxplots} to assess the distributions of HDs (calculated after post-processing). It can be seen that the HDs for the anomaly-aware method are much smaller overall.

\subsection{Anomaly-aware segmentation on OAI AKOA dataset with transfer learning}
\label{sec:results-transfer}

\begin{figure*}[!t]
\centering
\includegraphics[scale=.54]{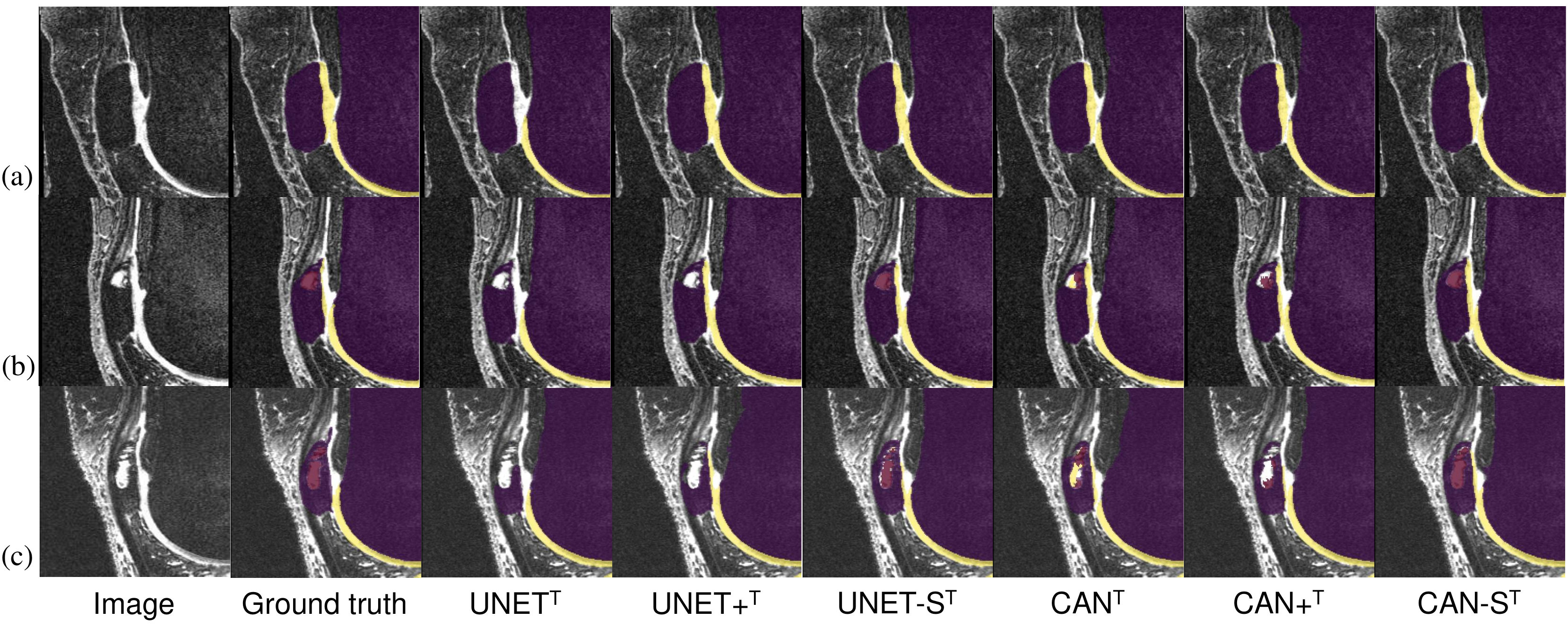}
\caption{Example outputs from the segmentation networks for images from the \textit{OAI AKOA} dataset, segmenting the patella and visible bone lesions in addition to the femur and tibia. Note that the images are zoomed in to view the patella more closely. The masks are overlaid on the input images (\textit{purple}: bones; \textit{yellow}: cartilages; \textit{red}: bone lesions). (a) When the images had no visible anomalies, all networks except $UNET^T$ produced good segmentation of the patella. The network $UNET^T$ failed to converge for the patellar cartilage label. (b,c) The anomaly-aware networks were able to detect and segment most of the visible lesions along with the anatomical structures on images that the other segmentation networks had difficulty with.}
\label{fig:segres-T}
\end{figure*}

Figure \ref{fig:segres-T} shows example outputs from the anomaly-aware segmentation networks $UNET$-$S^T$ and $CAN$-$S^T$ for the \textit{OAI AKOA} test sets and also the outputs from their baseline networks with transfer learning. As with Section \ref{sec:results-segmentation}, images with noticeable abnormalities tended to be more difficult to segment. For images with no visible anomalies, all networks except $UNET^T$ produced good segmentation of the patella (Figure \ref{fig:segres-T}(a)). The network $UNET^T$ did not converge for the patellar cartilage label for any of the training sets in 5-fold cross-validation. For images with visible abnormalities, $UNET$-$S^T$ and $CAN$-$S^T$ correctly detected most of the visible lesions and produced acceptable segmentations while the other networks occasionally failed (Figures \ref{fig:segres-T}(b) and \ref{fig:segres-T}(c)).

\begin{figure}[!t]
\centering
\includegraphics[scale=.43]{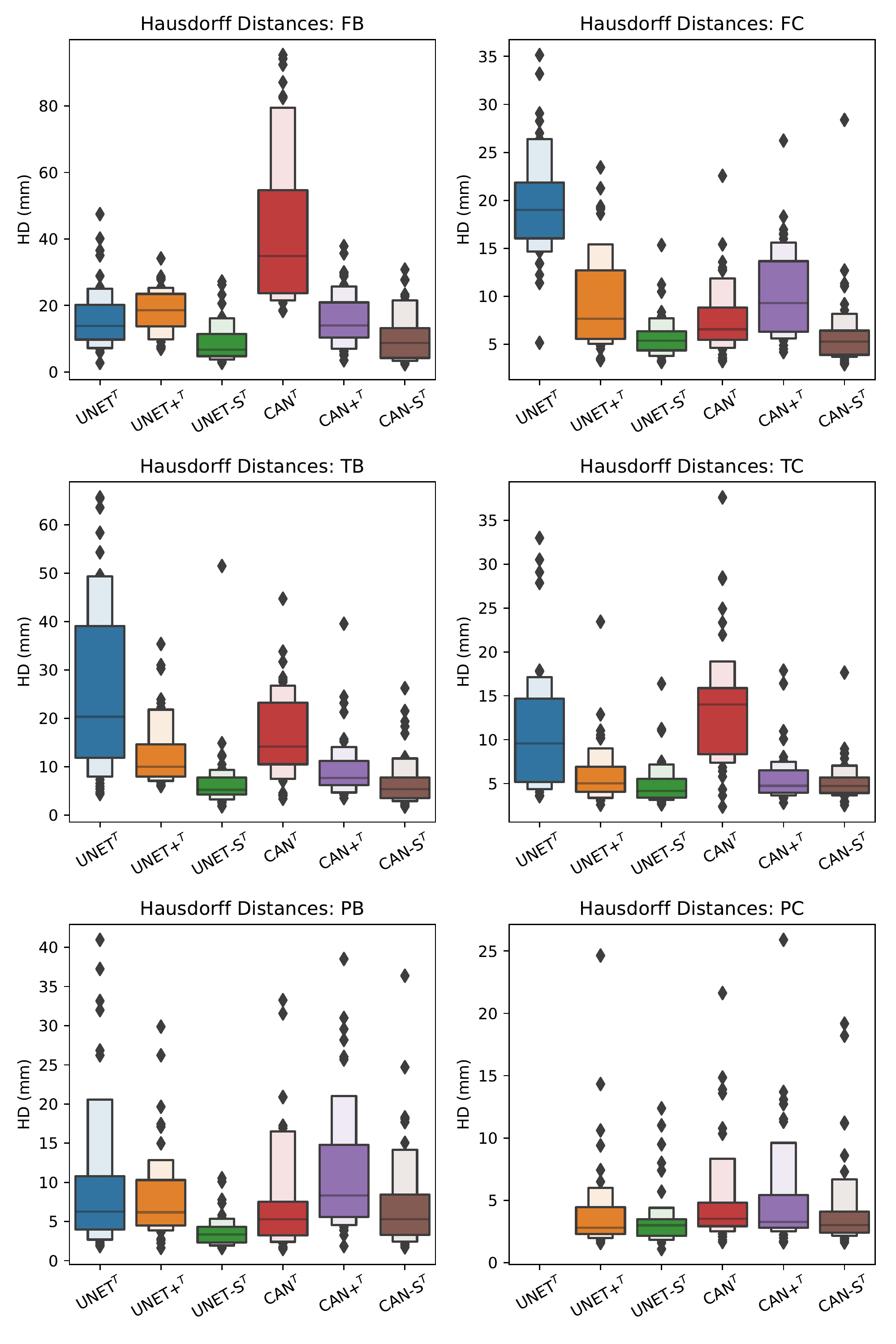}
\caption{Boxplots of Hausdorff distance (HD) values for the proposed anomaly-aware segmentation approach ($UNET$-$S^T$ and $CAN$-$S^T$) and baseline networks with transfer learning, evaluated on the \textit{OAI AKOA} dataset using 5-fold cross-validation. These HDs are results after post-processing. Note also that $UNET^T$ failed to converge for the PC label, so it was excluded from the plot.}
\label{fig:boxplots-T}
\end{figure}

Table \ref{tab:segres-T} shows the quantitative results for segmentation of the bones and cartilages. Again, there were statistically significant improvements in HDs for the anomaly-aware method compared to the baseline. In this case, the DSCs of the femoral bone and tibial bone volumes were also significantly improved. Note that the same post-processing method as in Section \ref{sec:results-segmentation} was applied here as well. The boxplots in Figure \ref{fig:boxplots-T} shows the distributions of HDs (calculated after post-processing). It can be observed that the HDs for the anomaly-aware method are much smaller than their baseline networks.

\begin{table*}[!t]
\caption{Mean DSC, ASD, and HD values for segmentation of the femoral, tibial, and patellar bone and cartilage volumes from the proposed anomaly-aware method ($UNET$-$S^T$ and $CAN$-$S^T$) with their baseline networks with transfer learning, evaluated using 5-fold cross-validation on the \textit{OAI AKOA} dataset ($N$ = 24$\times$2).}

\begin{center}
\begin{tabular}{|c|c||c|c|c||c|c|c|}
\hline
      \textbf{Class}& \textbf{Metric}& \textbf{UNET$^T$}& \textbf{UNET+$^T$}& \textbf{UNET-S$^T$}& \textbf{CAN$^T$}& \textbf{CAN+$^T$}& \textbf{CAN-S$^T$}\\
\hline
      ~& DSC (\%)& 96.9$\pm$1.63$^{\dag\ddag}$& 97.3$\pm$1.04$^{\dag\ddag}$& \textbf{\underline{98.4$\pm$0.41}}& 96.5$\pm$1.75$^{\dag\ddag}$& 97.0$\pm$1.30$^{\dag\ddag}$& \textbf{98.4$\pm$0.55}\\
      \multirow{2}{*}{\textbf{FB}}& ASD (mm)& 0.55$\pm$0.39$^{\dag\ddag}$& 0.47$\pm$0.26& \textbf{\underline{0.24$\pm$0.08}}& 0.97$\pm$0.91$^{\dag\ddag}$& 0.54$\pm$0.36$^{\dag\ddag}$& \textbf{0.26$\pm$0.12}\\
      ~& HD (mm)& 16.36$\pm$9.49$^\dag$& 18.21$\pm$6.56$^{\dag\ddag}$& \textbf{\underline{8.96$\pm$6.08}}& 41.78$\pm$22.94$^{\dag\ddag}$& 15.84$\pm$8.06$^\dag$& \textbf{10.24$\pm$7.62}\\
      ~& HD$^*$ (mm)& 41.16$\pm$24.56$^{\dag\ddag}$& 34.00$\pm$15.91$^{\dag\ddag}$& \textbf{\underline{10.70$\pm$9.96}}& 87.78$\pm$5.74$^{\dag\ddag}$& 30.71$\pm$21.26$^{\dag\ddag}$& \textbf{16.28$\pm$16.02}\\
\hline
      ~& DSC (\%)& 85.3$\pm$3.13& 85.9$\pm$2.67& \textbf{\underline{86.6$\pm$2.48}}& 84.5$\pm$3.10$^{\dag\ddag}$& 85.0$\pm$3.22& \textbf{86.4$\pm$2.53}\\
      \multirow{2}{*}{\textbf{FC}}& ASD (mm)& 0.32$\pm$0.07$^{\dag\ddag}$& 0.29$\pm$0.05& \textbf{\underline{0.26$\pm$0.04}}& 0.34$\pm$0.10$^{\dag\ddag}$& 0.31$\pm$0.06$^{\dag\ddag}$& \textbf{0.26$\pm$0.04}\\
      ~& HD (mm)& 19.75$\pm$5.58$^{\dag\ddag}$& 9.43$\pm$5.17$^{\dag\ddag}$& \textbf{\underline{5.73$\pm$2.25}}& 7.64$\pm$3.66& 10.41$\pm$4.64$^{\dag\ddag}$& \textbf{6.09$\pm$3.96}\\
      ~& HD$^*$ (mm)& 29.98$\pm$16.98$^{\dag\ddag}$& 14.07$\pm$10.51$^\dag$& \textbf{\underline{5.73$\pm$2.25}}& 15.99$\pm$17.43$^{\dag\ddag}$& 20.52$\pm$15.33$^{\dag\ddag}$& \textbf{7.22$\pm$6.70}\\
\hline
      ~& DSC (\%)& 97.4$\pm$0.93$^{\dag\ddag}$& 97.6$\pm$0.73$^{\dag\ddag}$& \textbf{\underline{98.4$\pm$0.31}}& 97.4$\pm$1.04$^{\dag\ddag}$& 97.3$\pm$1.40$^{\dag\ddag}$& \textbf{98.3$\pm$0.31}\\
      \multirow{2}{*}{\textbf{TB}}& ASD (mm)& 0.51$\pm$0.27$^{\dag\ddag}$& 0.37$\pm$0.14& \textbf{\underline{0.24$\pm$0.10}}& 0.42$\pm$0.20$^{\dag\ddag}$& 0.46$\pm$0.58$^{\dag\ddag}$& \textbf{0.24$\pm$0.08}\\
      ~& HD (mm)& 26.50$\pm$17.89$^{\dag\ddag}$& 12.55$\pm$7.12& \textbf{\underline{6.98$\pm$7.13}}& 16.61$\pm$8.97$^{\dag\ddag}$& 9.64$\pm$6.48& \textbf{7.06$\pm$5.42}\\
      ~& HD$^*$ (mm)& 107.47$\pm$6.61$^{\dag\ddag}$& 46.55$\pm$23.83$^{\dag\ddag}$& \textbf{\underline{7.96$\pm$7.96}}& 99.87$\pm$21.99$^{\dag\ddag}$& 39.56$\pm$25.70$^{\dag\ddag}$& \textbf{15.94$\pm$17.27}\\
\hline
      ~& DSC (\%)& 84.5$\pm$4.24& \textbf{84.7$\pm$3.66}& \textbf{\underline{85.0$\pm$3.16}}& 83.2$\pm$4.55& 83.6$\pm$4.50& 84.7$\pm$3.44\\
      \multirow{2}{*}{\textbf{TC}}& ASD (mm)& 0.31$\pm$0.14& 0.30$\pm$0.11& \textbf{\underline{0.28$\pm$0.11}}& 0.43$\pm$0.74& 0.32$\pm$0.13& \textbf{0.29$\pm$0.12}\\
      ~& HD (mm)& 11.08$\pm$7.35$^{\dag\ddag}$& 6.07$\pm$3.47& \textbf{\underline{4.84$\pm$2.53}}& 13.59$\pm$6.91$^{\dag\ddag}$& 5.66$\pm$2.97& \textbf{5.19$\pm$2.36}\\
      ~& HD$^*$ (mm)& 30.21$\pm$23.06$^{\dag\ddag}$& 19.61$\pm$18.21$^{\dag\ddag}$& \textbf{\underline{4.84$\pm$2.53}}& 36.27$\pm$21.11$^{\dag\ddag}$& 12.97$\pm$17.77& \textbf{9.19$\pm$13.59}\\
\hline
      ~& DSC (\%)& 96.0$\pm$1.29& 96.2$\pm$0.86& \textbf{\underline{96.6$\pm$0.80}}& 96.0$\pm$1.48& 95.8$\pm$1.29$^\dag$& \textbf{96.3$\pm$1.02}\\
      \multirow{2}{*}{\textbf{PB}}& ASD (mm)& 0.33$\pm$0.12& \textbf{0.30$\pm$0.07}& \textbf{\underline{0.26$\pm$0.07}}& 0.41$\pm$0.59& 0.42$\pm$0.31& 0.32$\pm$0.18\\
      ~& HD (mm)& 10.22$\pm$9.78$^\dag$& 8.27$\pm$5.86$^\dag$& \textbf{\underline{3.69$\pm$1.97}}& 7.64$\pm$7.17& 11.86$\pm$8.50$^{\dag\ddag}$& \textbf{7.43$\pm$6.69}\\
      ~& HD$^*$ (mm)& 78.45$\pm$20.18$^{\dag\ddag}$& 75.03$\pm$19.35$^{\dag\ddag}$& \textbf{\underline{10.33$\pm$21.00$^\ddag$}}& 94.30$\pm$27.55$^{\dag\ddag}$& 72.00$\pm$26.59$^{\dag\ddag}$& \textbf{29.04$\pm$34.58$^\dag$}\\
\hline
      ~& DSC (\%)& 0.0$\pm$0.00$^\S$& \textbf{85.1$\pm$5.48}& \textbf{\underline{85.7$\pm$5.02}}& 83.9$\pm$6.02& 84.5$\pm$5.43& 84.5$\pm$6.15\\
      \multirow{2}{*}{\textbf{PC}}& ASD (mm)& N/A$^\S$& \textbf{0.29$\pm$0.11}& \textbf{\underline{0.27$\pm$0.08}}& 0.33$\pm$0.13& 0.30$\pm$0.12& 0.31$\pm$0.12\\
      ~& HD (mm)& N/A$^\S$& \textbf{4.20$\pm$3.90}& \textbf{\underline{3.49$\pm$2.36}}& 4.93$\pm$3.99& 5.13$\pm$4.44& 4.29$\pm$3.70\\
      ~& HD$^*$ (mm)& N/A$^\S$& 24.23$\pm$30.61$^{\dag\ddag}$& \textbf{7.54$\pm$16.23}& 38.96$\pm$36.00$^{\dag\ddag}$& 28.89$\pm$33.77$^{\dag\ddag}$& \textbf{\underline{5.98$\pm$12.17}}\\
\hline
\multicolumn{8}{@{}l}{\small Bold with underline represents the best value within each metric. Bold without underline represents the second best value.}\\
\multicolumn{8}{@{}l}{\small HD$^*$ refers to Hausdoff distances before post-processing. All other metrics refer to results after post-processing.}\\
\multicolumn{8}{@{}l}{\small $^\dag$ represents significant difference (p-value $<$ 0.05) compared to UNET-S$^T$ with Tukey's HSD test.}\\
\multicolumn{8}{@{}l}{\small $^\ddag$ represents significant difference (p-value $<$ 0.05) compared to CAN-S$^T$ with Tukey's HSD test.}\\
\multicolumn{8}{@{}l}{\small $^\S$ UNET$^T$ failed to converge, so it was excluded from Tukey's HSD test (for the PC label only).}
\end{tabular}
\end{center}

\label{tab:segres-T}
\end{table*}

Performance for segmentation of bone lesions was less straightforward to evaluate because (a) there are a variable number of lesions (some images have none while some images have many) and (b) some lesions are too small to be reliably detected. In this study, we assumed ``bone-wise" lesion detection for simplicity, in which each bone was classified as either a positive or a negative case:
\begin{itemize}
\item For positive cases (bone has a lesion), a prediction was considered to be ``true positive" if the segmentation DSC $\ge$ 5\% and ``false negative" if DSC $<$ 5\%.
\item For negative cases (bone has no lesion), a prediction was considered to be ``true negative" if the network successfully predicted that there is no lesion in the bone and ``false positive" if the network predicted that there is one.
\end{itemize}
Note there are 3 bones (i.e. 3 ``cases") in each image: femur, tibia, and patella. In the 48 images from the \textit{OAI AKOA} dataset, there were 76 positive cases and 68 negative cases in total.

Using these criteria, sensitivity (true positive rate; TPR) was $>$ 90\% for most models while specificity (true negative rate; TNR) was $<$ 50\% (Table \ref{tab:ROC}). The specificity is low mainly due to the presence of subtle lesions in many images. See Figure \ref{fig:FP-FN} in the Appendix showing example images with very small lesions that often resulted in ``false positives" or ``false negatives". In most cases, the neural networks were highly sensitive and detected those small lesions, resulting in high false positive rates. However, such lesions are not consistently detected by human observers either and often are not clinically important. Therefore, we applied various size thresholds to the lesion masks to see if the specificity is higher for larger lesions. Using progressively increasing thresholds up to 6.0 mm$^3$, isolated lesions smaller than the threshold were removed from the output masks, and the same post-processing was applied to the manual masks for consistency. See \ref{sec:post-processing} for details on the post-processing for bone lesion detection.

Table \ref{tab:ROC} shows the bone lesion detection and segmentation performance of the different models in terms of accuracy and mean DSC. By progressively increasing the threshold, sensitivity was decreased but specificity was increased. The highest accuracy achieved with the post-processing was 84.7\% using $CAN$-$S^T$. The highest mean DSC (segmentation performance for positive cases only) was 53.6\% using $UNET$-$S^T$. The area under the receiver operating characteristic curve (AUC) was 0.896 for $CAN$-$S^T$ and 0.892 for $UNET$-$S^T$, both of which were higher than their baselines. Note in Table \ref{tab:ROC} that $UNET^T$ did not converge for the patellar lesion label. Its specificity is paradoxically high because it did not detect any patellar lesion, but the accuracy is still low due to low sensitivity. Another notable finding is that while $UNET$-$S^T$ performed the best in the segmentation tasks overall (Tables \ref{tab:segres} and \ref{tab:segres-T}), $CAN$-$S^T$ performed the best in the bone lesion detection task (Table \ref{tab:ROC}).

\section{Discussion}
\label{sec:discussion}

\subsection{Anomaly detection}
\label{sec:discussion-detection}

The current study presented a method to apply 3D U-Net-based CNNs for visual anomaly detection in volumetric medical images. There were two steps to this end, involving two separate networks $G$ and $A$. Network $G$ (the first step) could be sufficient if you already have the appropriate segmentation masks and you are only interested in highlighting anomalies in the images. However, segmentation masks are initially unavailable in most cases, so network $A$ (the second step) was added to carry out the same task without having to obtain segmentation masks first. The last column of Figure \ref{fig:recon} shows that the anomaly detection network $A$ can visually highlight anomalies on MR images of the knee.

These autoencoder-based networks also reconstruct the images with most of the visible lesions removed and therefore, they can theoretically be used to show what the images likely looked like if the bones had no lesions. The main limitation for this is that the quality of the images reconstructed from the current networks are rather poor. This is complex because it is actually easy for convolutional autoencoders to reconstruct images with minimal reconstruction error, but for anomaly detection, the reconstruction needs to be lossy since the anomalies should be removed, which means we might need to compromise on image quality. Nevertheless, a useful future work would include improving the model to make the images look more anatomically realistic. A possible approach would be to modify the image compressor, for example by using a more sophisticated model such as a conditional encoder or vector quantization \citep{vqvae}.

\subsection{Anomaly-aware segmentation}
\label{sec:discussion-segmentation}

This study also presented a generalized method for CNNs in which the information from the anomaly detectors is utilized to improve image segmentation. Although the main purpose was to compare the segmentation performance of the anomaly-aware networks to that of the baseline networks, there were also some incidental findings from the experiments such as the effect of deep supervision and the difference between U-Net and CAN.

First of all, it was demonstrated that the anomaly-aware mechanism is capable of improving the segmentation of overall bone volume on MR images of osteoarthritic knees with visible anomalies. The networks $UNET$ and $CAN$ were already capable of achieving high DSCs, but it was difficult to achieve good HDs without any shape regularization or post-processing. While deep supervision ($UNET+$ and $CAN+$) helped improve convergence and overall performance, it was found that the anomaly-aware method ($UNET$-$S$ and $CAN$-$S$) was able to further improve these for both U-Net and CAN. The anomaly-aware attention mechanism provided a substantial improvement in HDs in addition to a visible improvement in the quality of segmentation (Figures \ref{fig:segres}--\ref{fig:boxplots-T}). Although none of the CNNs were completely error-proof, as can be seen with the few outliers in HDs (Figures \ref{fig:boxplots} and \ref{fig:boxplots-T}), the significant drop in mean HDs with the anomaly-aware models indicates that these models make errors much less frequently. This is probably because the additional information provided to the network guided the network to focus its attention to the ROI, and the model was less vulnerable to anomalies and noise in the images.

The differences between HD$^*$s before post-processing and HDs after post-processing (Tables \ref{tab:segres} and \ref{tab:segres-T}) are also notable. The post-processing decreased HDs for most images for the non-anomaly-aware models, whereas it had an effect only for some images for the anomaly-aware models. Indeed, the post-processing made no difference in HDs for 94\% of all images (524/555) for $UNET$-$S$ and for 84\% (464/555) for $CAN$-$S$. In contrast, the post-processing made no difference in HDs for no images (0/555) for $UNET$ and $CAN$ and for 20\% (110/555) and 10\% (58/555), respectively, for $UNET+$ and $CAN+$. The inference time for the CNNs (without post-processing) was about 1 second per image, with the post-processing taking additional 4 seconds per image.

An additional finding from Table \ref{tab:segres} is that the U-Net-based models performed slightly better than the CAN-based models in the segmentation of femur and tibia on the \textit{OAI ZIB} dataset. However, $UNET^T$ failed to learn the patellar cartilage and patellar lesion labels on the \textit{OAI AKOA} dataset (Tables \ref{tab:segres-T} and \ref{tab:ROC}) despite using weighted Dice loss. Some possible reasons for the difficulty in learning these labels include (a) not enough training images, (b) highly imbalanced classes, (c) spatial heterogeneity, and (d) larger proportional impact of BMLs in the patella. An interesting observation is that the other models were still able to converge for all segmentation classes. Since the only difference between $UNET^T$ and $UNET+^T$ was deep supervision, it can be inferred that the deep supervision helped stabilize convergence. The more stable convergence behavior in $CAN^T$ compared to $UNET^T$ was initially unexpected but explainable; two possible explanations are as follows:
\begin{enumerate}
\item The CAN-based models are relatively ``simple", having a much less number of parameters than the U-Net-based models. For example, $CAN$ has 852,565 learnable parameters while $UNET$ has 5,887,765 learnable parameters. Having many parameters is not necessarily helpful because it can increase the chance of overfitting and the model might converge slower due to the model complexity.
\item Since the CAN-based models used minimal downsampling, they might have been better than $UNET^T$ at detecting very small structures such as the patellar cartilage and lesions. The U-Net-based models, on the other hand, rely on downsampling for aggregating multi-scale contextual information, which can be advantageous when segmenting larger anatomical structures but less desirable when having to analyze smaller structures. This is also consistent with the finding that although $UNET$-$S^T$ performed the best as a segmentation method, $CAN$-$S^T$ gave the best performance when used as a bone lesion detection method (Table \ref{tab:ROC}). Indeed, all CAN-based methods performed better than the U-Net-based methods on average in terms of accuracy and AUC.
\end{enumerate}

\begin{table*}[!t]
\caption{Bone lesion detection and segmentation performance on the \textit{OAI AKOA} dataset in terms of accuracy and mean DSC. Here, Acc. refers to the accuracy with no post-processing while $\lceil$Acc.$\rceil$ refers to the highest accuracy achieved with post-processing. Both are reported with the corresponding sensitivity (TPR) and specificity (TNR). $\lceil$DSC$\rceil$ is the highest mean DSC (averaged over all bone lesions) achieved with post-processing. AUC is the area under the receiver operating characteristic (ROC) curve. Note that $UNET^T$ failed to converge for the patellar lesion label. Results for each bone can be found in the supplementary material.}

\begin{center}
\begin{tabular}{|l|c|c|c|c|}
\hline
      \textbf{Model}& \textbf{Acc.} (TPR, TNR)& \textbf{$\lceil$Acc.$\rceil$} (TPR, TNR)& \textbf{$\lceil$DSC$\rceil$}& \textbf{AUC}\\
\hline
      \textbf{UNET$^T$}& 0.597 (0.632, 0.559)& 0.729 (0.592, 0.882)& 0.349& 0.713\\
      \textbf{UNET+$^T$}& 0.563 (0.921, 0.162)& 0.764 (0.750, 0.779)& \textbf{0.501}& 0.795\\
      \textbf{UNET-S$^T$} (Ours)& \textbf{0.681} (0.947, 0.382)& 0.819 (0.803, 0.838)& \textbf{\underline{0.536}}& \textbf{0.892}\\
\hline
      \textbf{CAN$^T$}& \textbf{0.681} (0.921, 0.412)& \textbf{0.833} (0.750, 0.926)& 0.462& 0.871\\
      \textbf{CAN+$^T$}& 0.611 (0.934, 0.250)& 0.806 (0.697, 0.926)& 0.464& 0.874\\
      \textbf{CAN-S$^T$} (Ours)& \textbf{\underline{0.694}} (0.921, 0.441)& \textbf{\underline{0.847}} (0.855, 0.838)& 0.493& \textbf{\underline{0.896}}\\
\hline
\multicolumn{5}{@{}l}{\small Bold with underline represents the best value within each metric.}\\
\multicolumn{5}{@{}l}{\small Bold without underline represents the second best value.}
\end{tabular}
\end{center}

\label{tab:ROC}
\end{table*}

Lastly, it can also be noted that the models with deep supervision ($UNET+^T$ and $CAN+^T$) did not outperform those without deep supervision ($UNET^T$ and $CAN^T$) in terms of detection accuracy; the models with deep supervision had higher sensitivity but also lower specificity, so the overall accuracy was not improved (Table \ref{tab:ROC}). The benefit of the anomaly-aware method is more apparent since the two models ($UNET$-$S^T$ and $CAN$-$S^T$) performed better in terms of accuracy, mean DSC, as well as AUC when compared to their baselines. The anomaly-aware models were able to maintain both high sensitivity and specificity. This is likely because these models already had information about where the lesions are likely to be whereas the baseline networks had to learn the information from scratch.

According to these findings, our proposed method is expected to be helpful in the analysis of medical images with visible anomalies, but the main limitation is that the current pipeline is rather involved. In Figure \ref{fig:pipeline}, it can be noted that the models $A$ and $S$ actually form a linear pipeline. Therefore, future work could investigate combining $A$ and $S$ into a single multi-task model to perform anomaly detection and segmentation simultaneously. In addition, since we can detect and segment pathologies, a future work may also include another downstream task such as classification of osteoarthritis grades. Osteophytes are also a major feature of osteoarthritis, but detection of osteophytes was not included in the current work because manual segmentation of osteophytes was challenging. A method to evaluate osteophyte detection may help with the development of the automated classification of osteoarthritis grades. Finally, since the anomaly-aware approach is a generalized method for CNNs, one could look into combining it with another CNN-based method such as nnU-Net \citep{isensee2021nnu} to further enhance the model.

\section{Conclusion}
\label{sec:conclusion}

In summary, this work demonstrated how simple U-Net-like neural networks can be used for detecting bone lesions in knee MR images through reconstruction via inpainting. Moreover, it showed how the detected anomalies can be further utilized for downstream tasks such as segmentation. The anomaly-aware networks gave a better performance on average than their baseline networks in the segmentation tasks as well as in the detection of bone lesions. The stable convergence behavior and performance with the new labels in the \textit{OAI ZIB--UQ} and \textit{OAI AKOA} datasets are promising and suggest that the proposed method has an advantage when there are relatively few training images and/or the classes are highly imbalanced. It is hoped that future works will show additional improvements and further applications of the anomaly detection and anomaly-aware segmentation models in medical imaging.

\section*{Acknowledgments}
%Acknowledgments should be inserted at the end of the paper, before the references, not as a footnote to the title. Use the unnumbered Acknowledgements Head style for the Acknowledgments heading.
We would like to thank Dr Jessica Bugeja for curating the \textit{OAI AKOA} dataset.

\appendix

\section{Implementation details}
\label{sec:implementation}

All of the neural networks in this study were implemented using Tensorflow \citep{tensorflow2015-whitepaper} version 2.4 with Keras API (\url{http://tensorflow.org/guide/keras}) and were trained on a high-performance computer with NVIDIA Tesla V100-SXM2-32GB.

The two anomaly detection networks $G$ and $A$ (Sections \ref{sec:methods-G} and \ref{sec:methods-A}; Figure \ref{fig:networks}) were both based on 3D U-Net \citep{cciccek20163d}, consisting of a contracting path (encoder) and an expansive path (decoder) with skip connections. Overall, the networks had 5 levels of resolution with 4 progressive downsamplings followed by 4 progressive upsamplings using strided convolutions. In the contracting path, there were two convolution blocks at each level, where each convolution block was a 3D convolution layer with a kernel size of 3 $\times$ 3 $\times$ 3 followed by an instance normalization \citep{ulyanov2016instance} and leaky rectified linear unit (ReLU) activation function with a negative slope coefficient of 0.1. Instance normalization was used because the large image size only allowed for a batch size of 1. The number of feature maps was progressively increased as the resolution was decreased.

\begin{figure*}[!t]
\centering
(a)\includegraphics[scale=.5]{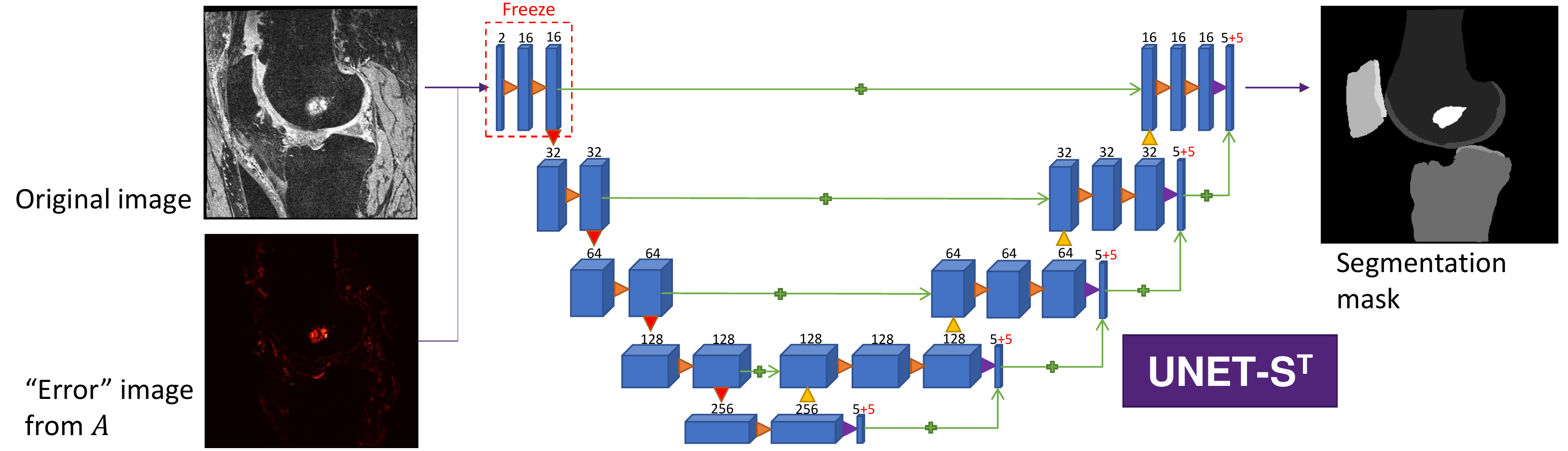}\\
(b)\includegraphics[scale=.5]{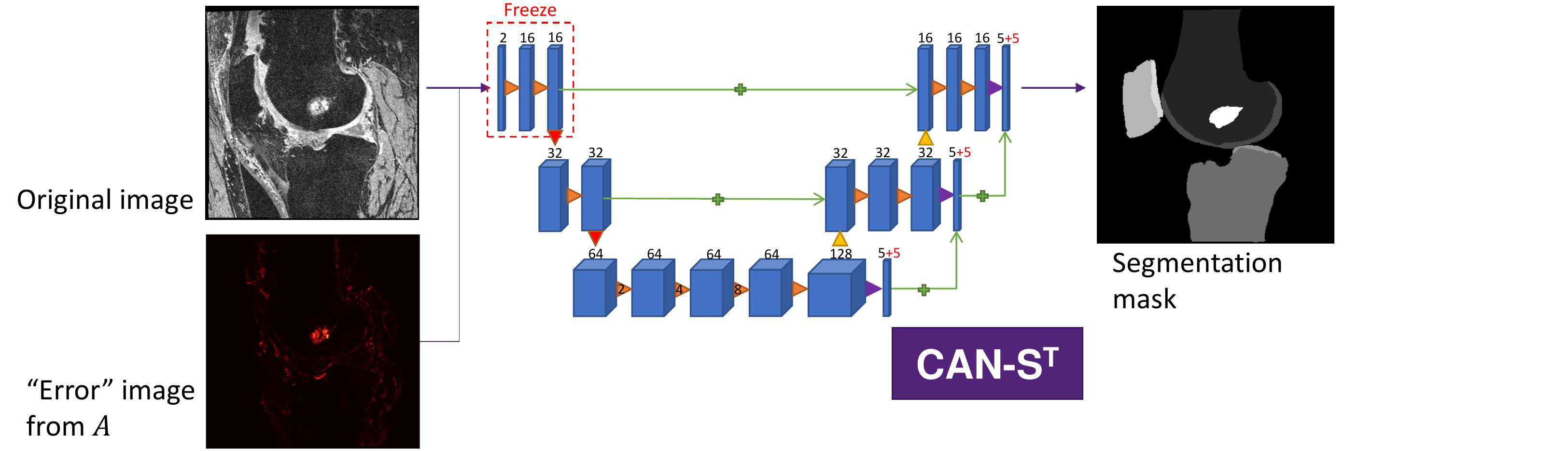}\\
\caption{The anomaly-aware segmentation network $S^T$ for transfer learning based on (a) 3D U-Net and (b) 3D CAN. The network $S^T$ is a slight modification from $S$ (Figure \ref{fig:segnet}) where 5 more channels were added to the output layer. During training, the first two convolution blocks were frozen.}
\label{fig:segnet-T}
\end{figure*}

The image compressor $C$ was a very small network, again with 4 progressive downsamplings, but with the number of feature maps progressively decreasing as the resolution was decreased. To create a bottleneck, the downsampled image was flattened, and then a dense layer with 100 units was applied. Then, after another dense layer with 5760 units, the flattened image was reshaped into a 3D image of 10 $\times$ 24 $\times$ 24 to be added to the decoder part of the U-Net.

In the expansive path of U-Net, the resolution was progressively recovered using transposed convolutions, and the feature maps from the encoder at the corresponding level were transferred using element-wise summation. The combined feature maps then passed through two more convolution blocks before another transposed convolution for upsampling. After reaching the original image resolution, the final output convolution layer with a kernel size of 1 $\times$ 1 $\times$ 1 and sigmoid activation function was applied. The MR images were Z-normalized (centered to 0 mean, unit standard deviation), clipped at [-5, 5], and subsequently rescaled to [0, 1] for network input. The models were trained using the Adam optimizer \citep{kingma2014adam} with a learning rate of 0.0005 for 50 epochs for each training set.

For the downstream segmentation task (Section \ref{sec:methods-S}; Figure \ref{fig:segnet}), two different types of CNNs were used: 3D U-Net and 3D CAN. The architecture of the U-Net-based models ($UNET$, $UNET+$ and $UNET$-$S$) was similar to $G$ and $A$, except that the final activation function was softmax and that $UNET+$ and $UNET$-$S$ were modified with deep supervision.

The CAN-based models ($CAN$, $CAN+$ and $CAN$-$S$) had 3 levels of resolution, where the first two levels were basically the same as those of U-Net. After two downsamplings, the CAN module was applied instead of further downsampling. The CAN module consisted of convolution blocks with progressively increasing dilation rates and then a non-dilated convolution block as the final block of the CAN module. A convolution block was almost the same as that defined above for U-Net except its dilation rates and kernel initializer. While non-dilated convolution layers were initialized using the Glorot uniform initializer \citep{pmlr-v9-glorot10a}, dilated convolutions were initialized with the identity initializer which was found to be more effective for context aggregation \citep{DBLP:journals/corr/YuK15}. After the CAN module, the resolution was recovered with transposed convolutions and skip connections as with U-Net, and then the final output layer was applied.

For the segmentation networks, the ``mixed precision" policy in Keras API was used to enable the use of a larger number of feature maps in 3D CNNs. Mixed precision refers the use of 16-bit floating-point type in parts of the model during training to make it use less memory. The output layer (softmax layer) was kept in the 32-bit type for numerical stability. The MR images were Z-normalized for network input, and the models were trained using the Adam optimizer with a learning rate of 0.0005 for 50 epochs for each training set.

A few modifications were made to all segmentation networks for transfer learning (Section \ref{sec:transfer}; Figure \ref{fig:segnet-T}). Firstly, 5 more channels were added to the output layer to segment the additional classes. In addition, since the initial layers in CNNs tend to extract general image features, weights for the first two convolution blocks were frozen during training, and the rest of the network was trained with a slightly lower learning rate of 0.00025 for 50 epochs for each training set. Due to the limited size of the datasets, online data augmentation was applied to reduce overfitting. The data augmentation consisted of random scaling (scaling factor within [0.9, 1.1]), random rotation (angle within [-10, 10] degrees in a random direction), and random translation (within [-10, 10] pixels in each direction).

\section{Evaluation metrics}
\label{sec:evaluation}

Segmentation performance was evaluated using Dice similarity coefficient (DSC), which is defined as:
\begin{equation}\label{eq:DSC}
DSC=\frac{2|B\cap A|}{|B|+|A|}.
\end{equation}
Here, $A$ and $B$ denote the set of positive voxels in the ground truth segmentation map and the predicted segmentation map, respectively. DSC is a volumetric measure usually expressed as a percentage. Although it is useful for general assessment of the overall segmentation results, it provides limited sensitivity to errors on the boundaries of the segmentation if the segmented volume is large. We therefore also included surface distance measures for the anatomical structures, to evaluate segmentation errors on their boundaries. The average surface distance (ASD) and Hausdorff distance (HD; also known as maximum surface distance) are defined as:
\begin{equation}\label{eq:ASD}
ASD=\frac{1}{n_{\partial(A)}+n_{\partial(B)}}\left(\sum_{a\in\partial(A)}d(a,B)+\sum_{b\in\partial(B)}d(b,A)\right),
\end{equation}
\begin{equation}\label{eq:HD}
HD=\max\left(\max_{a\in\partial(A)}d(a,B),\max_{b\in\partial(B)}d(b,A)\right).
\end{equation}
Here, $\partial(\cdot)$ denotes the boundary of the segmentation set, $n_{\partial(\cdot)}$ denotes the number of voxels on the boundary $\partial(\cdot)$, and
\begin{equation}
d(p,Q)=\min_{q\in\partial(Q)}||p-q||_2
\end{equation}
is the minimum distance from point $p$ to the boundary of a segmentation set $Q$.

\section{Post-processing of CNN outputs}
\label{sec:post-processing}

For the bone and cartilage labels, random stray voxels in the CNN outputs were removed by extracting the largest component for each label and removing smaller components that are more than 50 voxels away from the largest component. Simply extracting the largest component only (i.e. removing \textit{all} smaller components) was invalid because sometimes that also removed correctly segmented voxels. For example, in some images with very severe cartilage degeneration, there might be smaller fragments of cartilages, so removing all smaller components sometimes makes the segmentation worse. Therefore, some allowance had to be made for the stray voxels that are relatively close to the main structures even though that means not all stray voxels could be removed. The post-processing still removed most of the very extremely stray voxels, significantly reducing mean HDs for the non-anomaly-aware models.

\begin{figure}[!t]
\centering
\includegraphics[scale=.54]{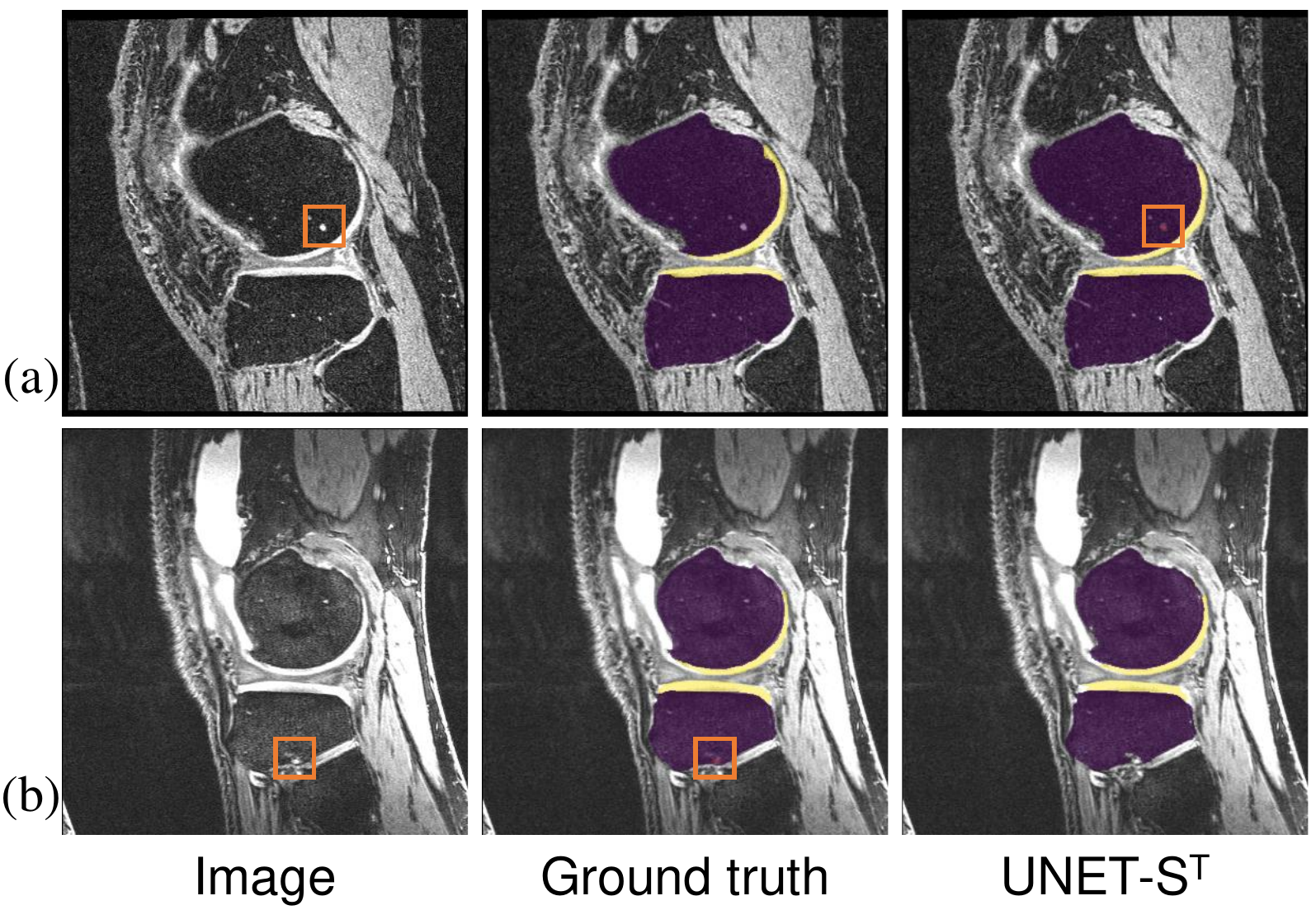}
\caption{Example of (a) ``false positive" and (b) ``false negative" case, which mostly occurred with very small lesions. The orange boxes highlight the small lesions.}
\label{fig:FP-FN}
\end{figure}

For the bone lesion labels, a different post-processing method was used since there are a variable number of lesions and many of the lesions are quite small in size. Figure \ref{fig:FP-FN} shows example images with very small lesions that often resulted in ``false positives" or ``false negatives", decreasing the specificity of the segmentation networks for the bone lesion detection task (Section \ref{sec:results-transfer}). To see if the specificity is higher for larger lesions, various size thresholds were applied to the lesion masks. Using progressively increasing size thresholds of 0.0 (no post-processing), 0.5, 1.0, ..., 6.0 mm$^3$, isolated lesions smaller than the threshold were removed from the output masks, with the same post-processing applied to the manual masks for consistency.

For $>$ 6.0 mm$^3$ size threshold, some positive cases changed to negative cases, paradoxically increasing the sensitivity of the detection method. (It means all positive cases had an isolated lesion at least 6.0 mm$^3$.) At 6.0 mm$^3$ threshold, the specificities of the models ranged from 68--88\%. In order to complete the receiver operating characteristic (ROC) curves, the size threshold was fixed at 6.0 mm$^3$, and then the specificity of the detection method was increased by increasing the softmax output threshold. The voxels with softmax output (probability) less than the threshold were removed from the output masks, using thresholds of 0.5, 0.6, 0.7, ..., 1.0.

The areas under the ROC curves (AUCs) along with the highest accuracies and the highest mean DSCs achieved with post-processing are reported in Table \ref{tab:ROC}.

%\section*{Supplementary Material}
%Supplementary material that may be helpful in the review process should be prepared and provided as a separate electronic file. That file can then be transformed into PDF format and submitted along with the manuscript and graphic files to the appropriate editorial office.

%%Harvard
\bibliographystyle{model2-names.bst}\biboptions{authoryear}
\bibliography{refs}

\end{document}